\documentclass[traditabstract]{aa}

\usepackage{graphicx}
\usepackage{amsmath}
\usepackage{color}
\usepackage{natbib}
\bibpunct{(}{)}{;}{a}{}{,}

\begin{document}

\title{Radial migration in a stellar galactic disc with thick components}
\titlerunning{Radial migration in a galactic disc with thick components}

\author{A. Halle        \inst{1}  
        \and
       P. Di Matteo        \inst{2}
       \and
       M. Haywood    \inst{2}
       \and
       F. Combes   \inst{3}
}

\offprints{A. Halle}

\institute{Max Planck Institut f\"ur Astrophysik, 
        Karl-Schwarzschild-Strasse 1, 
        D-85741 Garching bei M\"unchen, Germany
\and
Observatoire de Paris, GEPI, 
	5 place Jules Janssen
 	92190 Meudon, France
 	\and
 	 LERMA, Observatoire de Paris, Coll\`ege de France, CNRS, PSL Univ., UPMC, Sorbonne Univ., Paris, France\\
                \email{halle@mpa-garching.mpg.de}
}

\date{}

\abstract{We study how migration affects stars of a galaxy with a thin stellar disc and thicker stellar components. The simulated galaxy has a strong bar and lasting spiral arms. We find that the amplitude of the churning (change in angular momentum) is similar for thin and thick components, and of limited amplitude, and that stars of all components can be trapped at the corotation of the bar. At the exception of those stars trapped at the corotation, we find that stars that are far from their initial guiding radius are more likely so due to blurring rather than churning effects. We compare the simulation to orbits integration with a fixed gravitational potential rotating at a constant speed. In the latter case, stars trapped at corotation are churned periodically outside and inside the corotation radius, with a zero net average. However, as the bar speed of the simulated galaxy decreases and its corotation radius increases, stars trapped at corotation for several Gyrs can be churned outwards on average. We study the location of extreme migrators (stars experimenting the largest churning) and find that extreme migrators come from regions on the leading side of the effective potential local maxima.}


\maketitle

\section{Introduction} 

Observations, theoretical and/or numerical studies have shown that radial migration is of interest to understand some metallicity/chemistry observations such as the age-metallicty scatter in the solar neighbourhood \citep[e.g.][]{haywood08, schoenrich09}, stellar metallicity distributions at different radii in the Milky Way \citep{loebman16} or the upturn of mean stellar age in the outskirts of local galaxies \citep[e.g.][]{roskar08, bakos08, ruizlara17}. 

Stars guiding radius can change because of dynamical interactions during mergers, with clumpy structures in the discs or with non-axisymmetric patterns like bars or spiral arms \citep[e.g.][]{lyndenbell72, sellwood02}. A number of studies have focused on radial migration generated by resonance with such non-axisymmetric patterns \citep[e.g.][]{daniel15, minchev10, brunetti11, minchev11, minchev12, dimatteo13, grand12, grand14, kubryk13, roskar12, veraciro14, halle15}. This change of guiding radius is often named churning as opposed to blurring that is due to epicyclic oscillations around the guiding radius \citep{schoenrich09}. The churning can be oscillatory in the case of interaction with a lasting non-axisymmetric pattern such as long-lived spiral arms or long-lived bars \citep{binneytrem08, sellwood02, ceverino07}. This has led to consider that bars, that seem to be long-lived (even if they can be destroyed by gas infall \citep{bournaud02} and grow again), could not drive substantial radial migration \citep[e.g.][]{aumer16} because stars are periodically churned back and forth in a region around corotation. In contrast, stars corotating with transient spiral arms are able to change their angular momentum permanently when transient spiral vanish \citep{sellwood02}. However, the churning driven by a bar mixes significantly the stars around the corotation of the bar, and, if the bar has a non-constant speed, the portion of the disc affected by churning can shift towards the inside or outside of the disc, in case of an increase \citep[e.g.][]{halle15, ceverino07} or decrease of the bar-speed, respectively. 

Radial migration has been studied in some simulations including thick discs \citep[e.g.][]{solway12, aumer17b} or also in the context of its potential thickening effect on discs \citep[e.g.][]{schoenrich09, schoenrich17, kubryk14a, loebman11, minchevvert12, roskarvert13, veraciro14, veraciro16, veraciroaction16, grand16}. Some studies however suggest that, at least in the Milky Way, the characteristics of the thick disc do not seem to require any significant migration and can be well explained by formation from a disc rich in gas and turbulent \citep[e.g.][]{noguchi98,brook04,haywood13,lehnert14, haywood15}. It has been argued that stars migrating from the inner disc parts to outer regions should heat the outer regions because they come from higher velocity dispersion regions \citep{schoenrich09, roskarvert13}, but this may be balanced by the provenance bias of the migrating stars \citep{veraciro14} (see also \citet{minchevvert12, veraciro16}): the stars that are more likely to migrate are the stars remaining close to the disc mid-plane, hence stars with a globally low velocity dispersion. \citet{solway12} found the churning in a thick disc is only mildly more important than the churning of thin disc stars.    

In this work, we study how stars that remain trapped around the bar corotation can be globally churned outwards in the case of a slowing-down bar whose corotation radius increases with time, in a disc galaxy with thick components. We use a N-body simulation of an isolated disc galaxy with three disc components of different scale heights, embedded in a live dark-matter halo, and compare the radial migration to the case of a fixed gravitational potential rotating at a constant speed. 

Section~\ref{section-num-sim} presents the numerical simulation used in this work, its initial conditions and the dynamical evolution of the stellar disc and its non-axisymmetric patterns. Section~\ref{section-rad-migr} focuses on radial migration of the thin and thicker disc components with a comparison to the radial migration in the fixed potential rotating at a constant speed case. Section~\ref{section-extreme-migr} focuses on the extreme migrators churned outwards, comparing their orbits and spatial location to the fixed potential rotating at a constant speed case.

\section{Numerical simulation}
\label{section-num-sim}
\subsection{Initial conditions}
\label{subsec-ics}

\begin{table*}
\centering
\begin{tabular}{lccccc}
 & Mass [$M_{\odot}$] & $a$ [kpc] & $h$ [kpc] & $r_h$ [kpc]  & $N_{\rm particles}$\\ 
 \hline
Thin disc & $2.6 \, 10^{10}$ &  4.7 & 0.3 & & $1  \, 10^{7}$  \\ 
Intermediate disc & $1.5 \, 10^{10} $ & 2.3 & 0.6 &   & $6  \, 10^{6}$\\
Thick disc & $1.0 \, 10^{10}$ & 2.3 & 0.9 & & $4  \, 10^{6}$ \\
Dark matter halo & $1.6 \, 10^{11}$ & & & 10 & $5 \, 10^{6}$\\
\end{tabular}
\label{tab-sim}
\caption{Galaxy parameters}
\end{table*}

The simulated galaxy contains a stellar disc with three components of different vertical scale heights, and a dark matter halo. No stellar bulge is included. Masses of the different components, number of particles and parameters are shown in Table~\ref{tab-sim}.

The discs are called thin, intermediate and thick in reference to their relative average thickness. They have Miyamoto-Nagai density profiles: 
\begin{multline}
\rho(R,z)=\frac{h^2 M_g}{4\pi} \\ 
\times \frac{a R^2+\left( a+3 \sqrt{z^2+h^2}\right) \left( a+\sqrt{z^2+h^2} \right)^2  }{ \left( R^2+ \left( a+ \sqrt{z^2+h^2}\right)^2  \right)^ {\frac{5}{2}} \left(z^2+h^2 \right)^{\frac{3}{2}} },
\end{multline}
with height parameters $h_{\mathrm{thin}}<h_{\mathrm{inter}}<h_{\mathrm{thick}}$ and radial parameters $a_{\rm thin}$ and lower $a_{\rm inter}=a_{\rm thick}$, as detailed in Table~\ref{tab-sim}. The three disc components of different scale heights aim at representing a more realistic total disc than a case with only a thin and a thick component (this is suggested by some studies of stellar populations in the Milky Way indicating a continuous variation of disc properties with scale height \citep[e.g.][]{bovy12}). In the following analyses, we however often show results for the three individual components because the intermediate case is an interesting transition between the thinner and thicker components for the study of radial migration. 

The dark matter halo has a Plummer profile:
\begin{equation}
\rho(r)=\frac{3M}{4\pi r_h}\left( 1+\frac{r^2}{{r_h}^2}\right)^{-\frac{5}{2}} 
\end{equation}

\begin{figure}[h!]
\centering
\resizebox{\hsize}{!}{\includegraphics{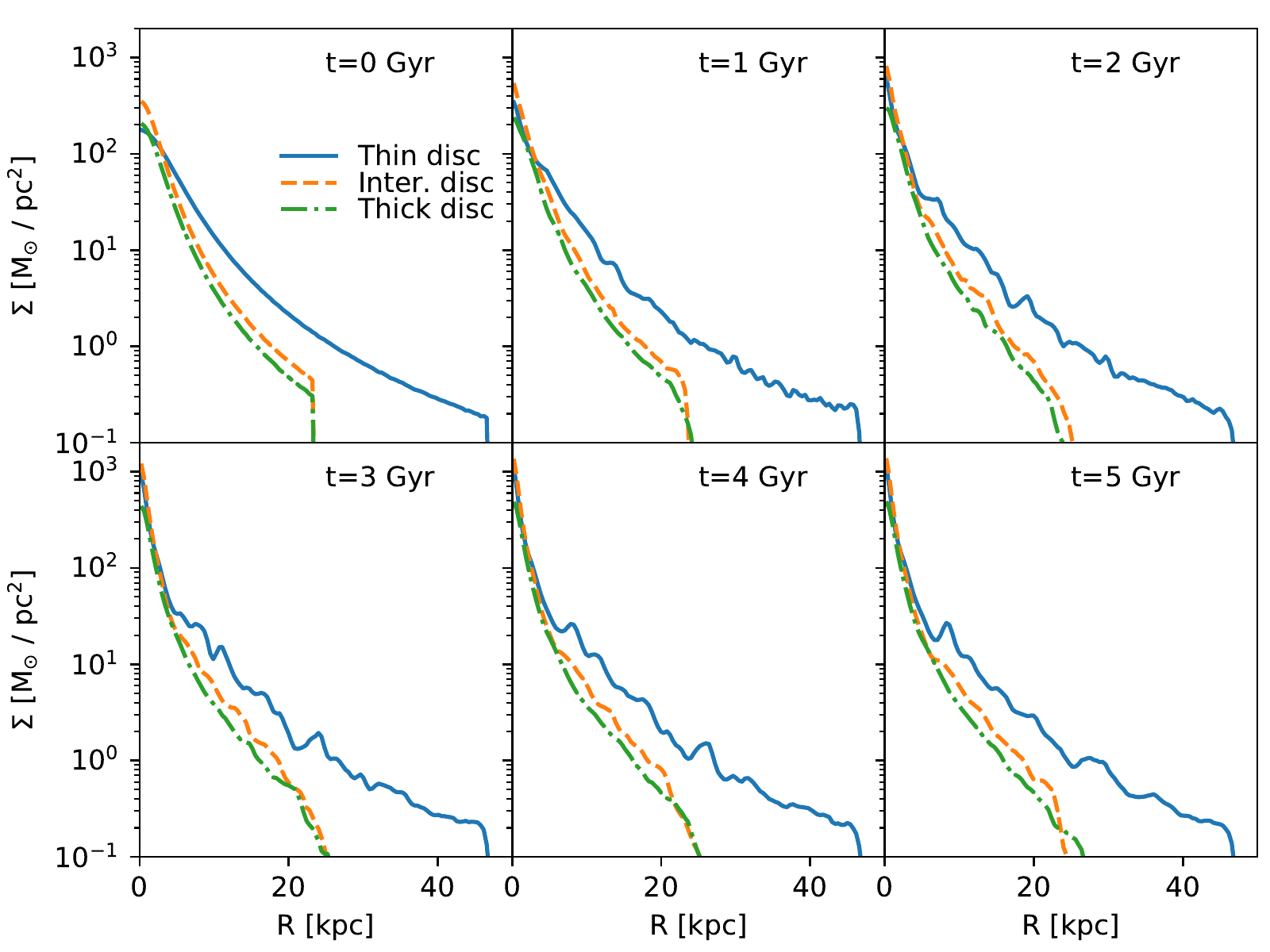} }
\caption{Surface density radial profiles of disc components at different times.}
\label{surfdens-fig}
\end{figure}

The initial conditions are set with an iterative method allowing the components to be dynamically relaxed. Radial migration due to resonance with non-axisymmetric patterns in isolated discs may sometimes be overestimated because of radial expansion of the stellar discs from unrelaxed initial conditions, which is avoided in this work. The low increase in radial extent of the discs is seen on Fig.~\ref{surfdens-fig}, showing the radial surface density profiles of the different disc components at initial and later times.  More details on the initial conditions are shown in Appendix~\ref{rotcurves-app}: contributions of the different components to the rotation curve are shown on Fig~\ref{vc-fig}, and the radial and vertical velocity dispersions radial profiles of the different disc components are shown on Fig.~\ref{dispr-fig} and Fig.~\ref{dispz-fig}.

\subsection{Dynamical evolution}

\begin{figure*}
\centering
\resizebox{\hsize}{!}{\includegraphics{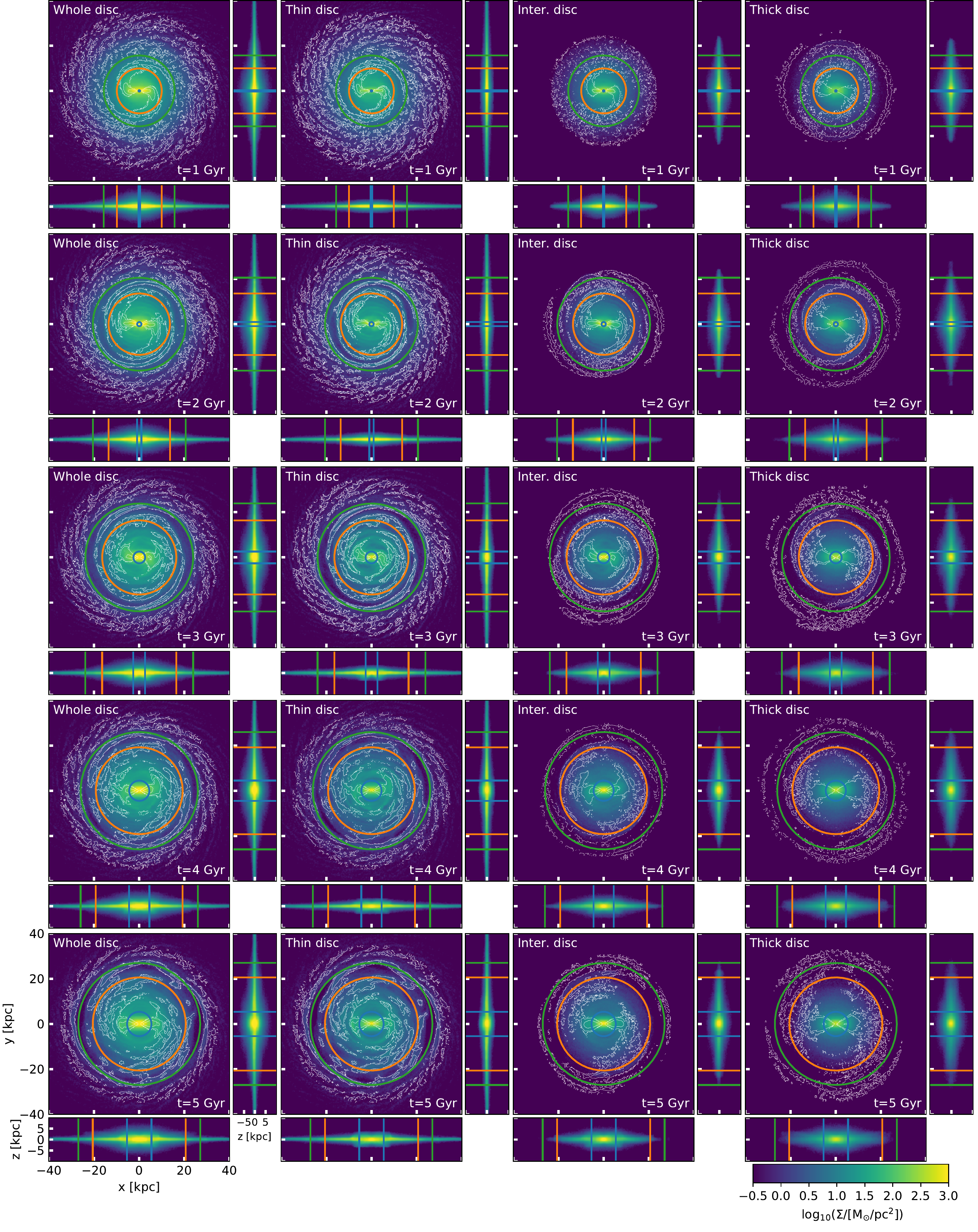}}
\caption{Face-on and edge-on surface density maps of disc components and the whole disc at different times. White contours: azimuthal overdensity (see text for details). Circles (and lines on edge-on views): ILR (blue), corotation (orange) and OLR (green) radii.}
\label{snaps-fig}
\end{figure*}

\begin{figure*}
\centering
\begin{tabular}{cc}
\includegraphics[width=7cm]{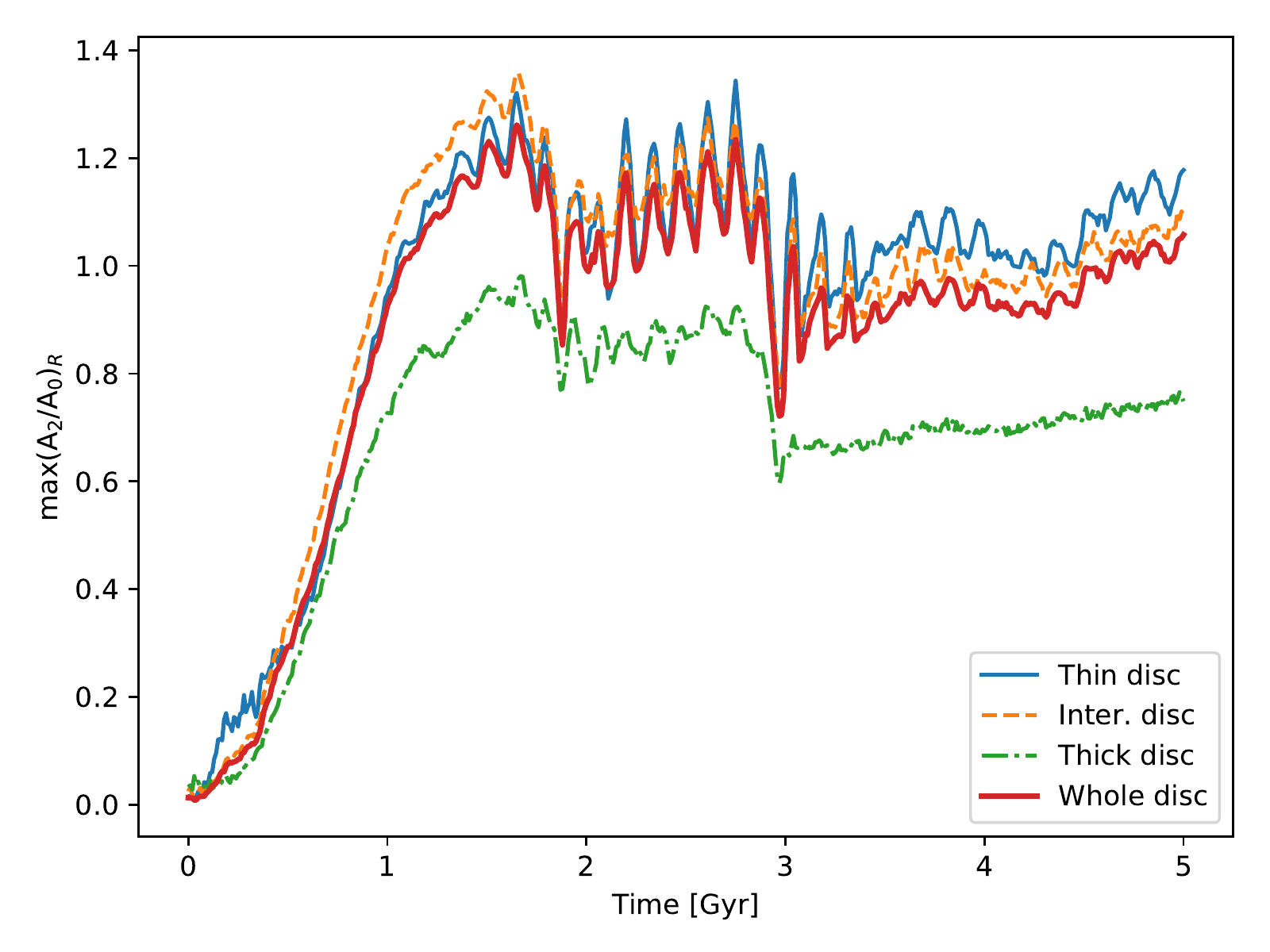} & \includegraphics[width=7cm]{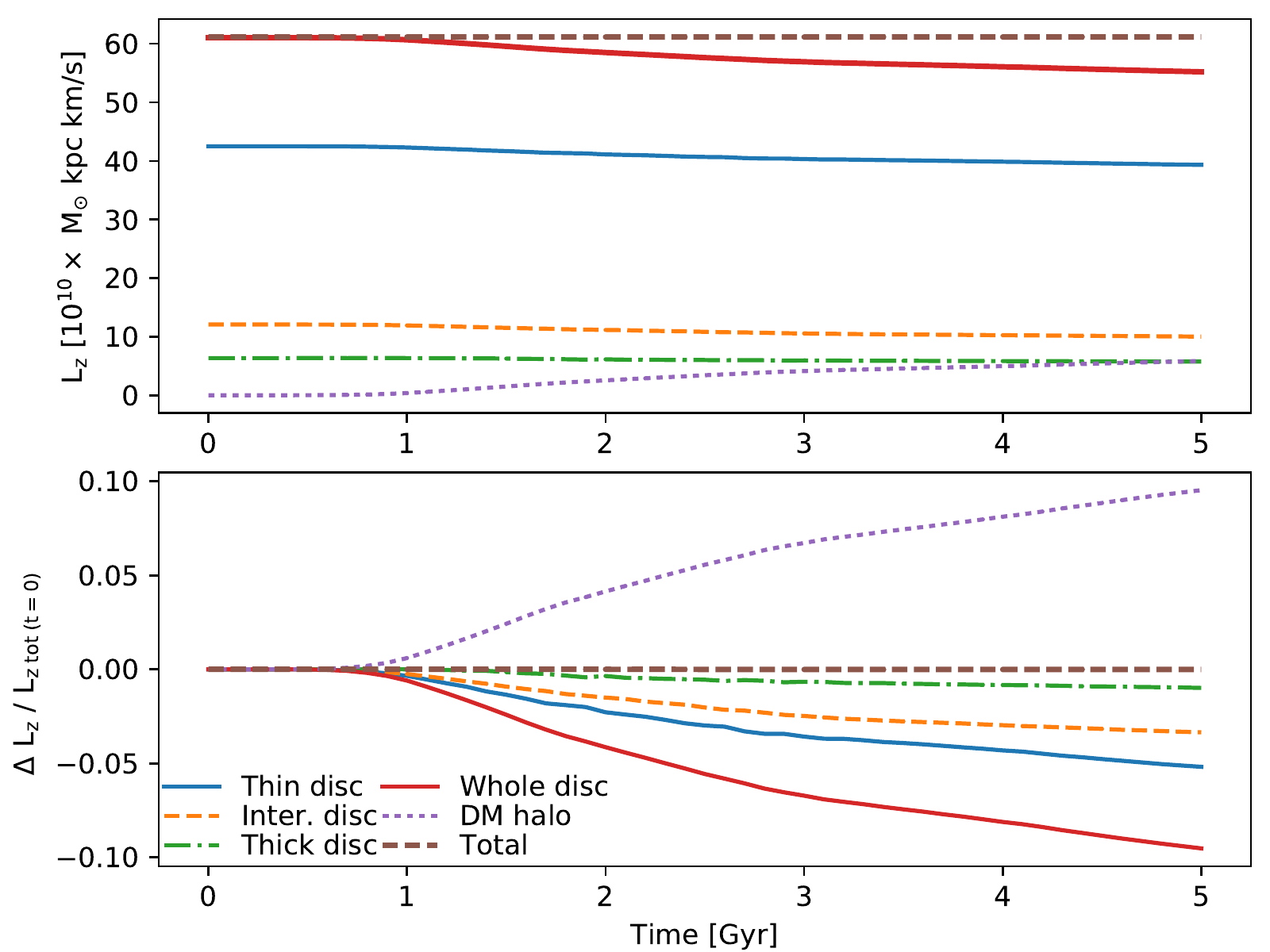}
\end{tabular}
\caption{Left:Bar strength evolution. Right: Vertical angular momentum evolution of the different galaxy components (top) and their variation relative to the total initial vertical angular momentum (bottom).}
\label{barst-fig}
\end{figure*}

\begin{figure*}
\centering
\resizebox{\hsize}{!}{\includegraphics{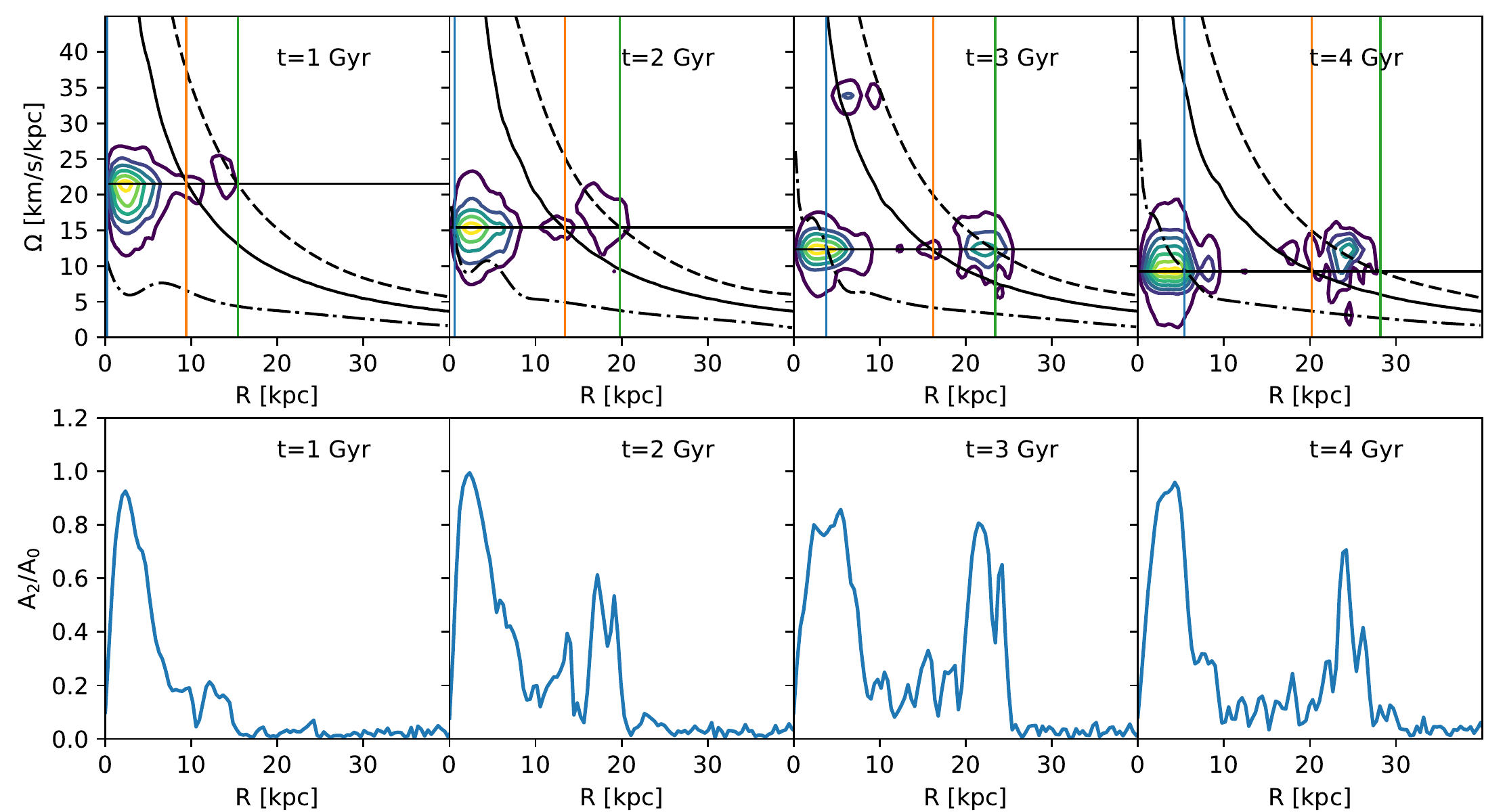} \includegraphics{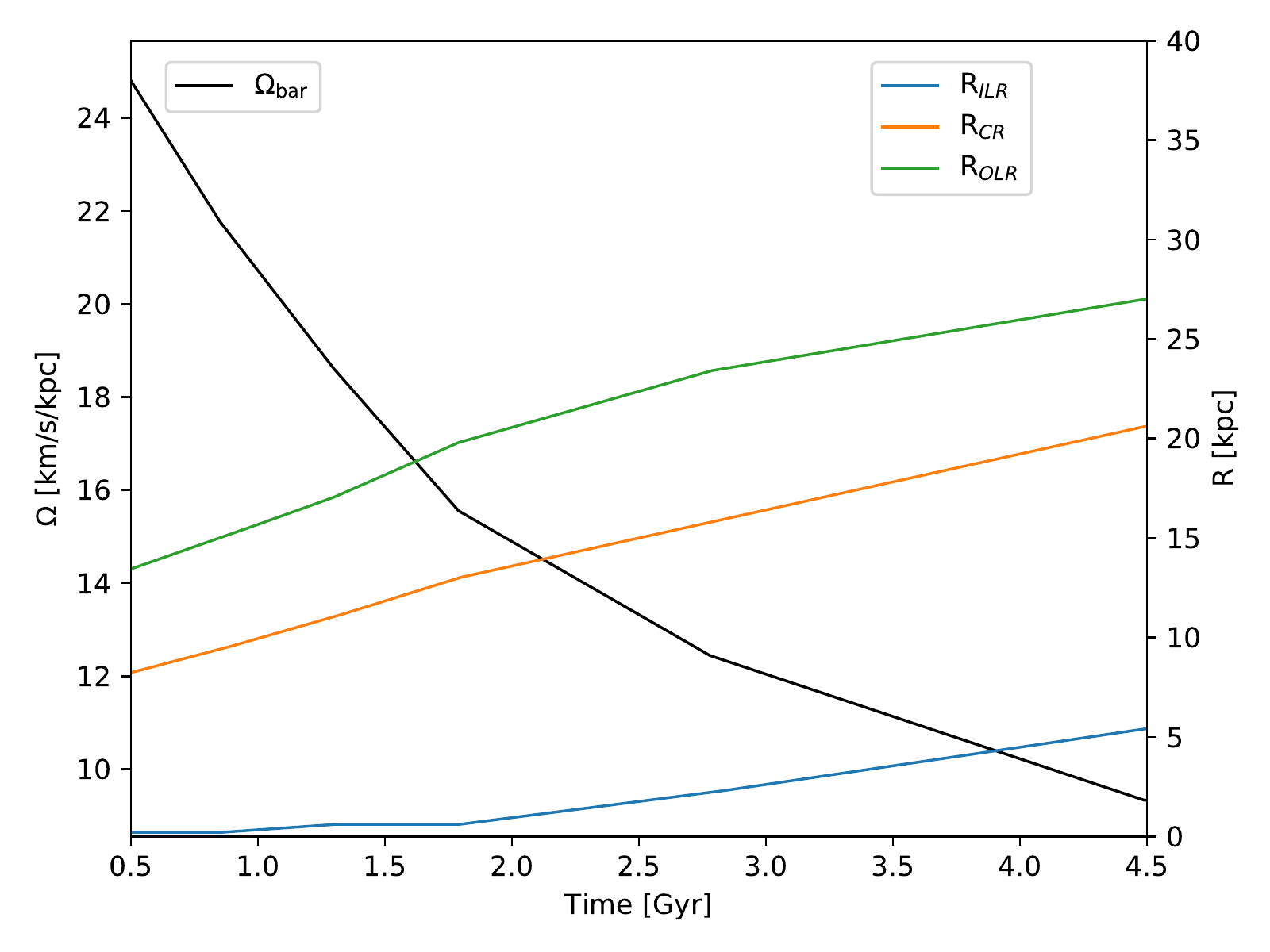} }

\caption{Left first row: Determination of resonances radii at different times from a Fourier analysis on the whole disc. Horizontal lines: pattern speed $\Omega_p$ obtained from the maximum of the spectrogram, solid curve: $\Omega(R)$, dashed curve: $\Omega(R)+ \dfrac{\kappa}{2}$, dash-dotted curve: $\Omega(R)- \dfrac{\kappa}{2}$. Vertical lines: radii of the ILR (blue), corotation (orange), and OLR (green). Left second row: Strength of $\pi$-periodic non axisymmetries of the whole disc as a function of radius at the same times as in the first row. Right plot: Resulting bar speed and resonance radii estimation as a function of time.}
\label{four-fig}
\end{figure*}

The simulated time-span is of 5~Gyrs. We use the Tree code of \citet{semelin02} with a softening length set to $50$~pc. Fig~\ref{snaps-fig} shows maps of the surface density of the whole disc and its three components after 1, 2, 3, 4 and 5~Gyr of evolution. Over-plotted in white are the positive isocontours of $\dfrac{\Sigma(R, \theta)- \Sigma(R)}{\Sigma(R)}$, the relative deviation of surface density at radius $R$ and azimuth $\theta$, $\Sigma (R, \theta)$, from the azimuthally averaged value $\Sigma (R)$.

The disc develops a bar and spiral features. The bar is the strongest non-axisymmetric perturbation at almost all times. Spiral arms have the same angular speed as the bar. Fig.~\ref{barst-fig} shows the time evolution of the bar strength as estimated from a Fourier decomposition of the surface density (from stars at $|z| < 500~pc$) of either each individual disc component or the whole components. Note that this simulation does not aim at reproducing the bar properties of the Milky Way. The bar in the thick disc is significantly weaker (as seen in \citep[e.g.][]{combes90, athanassoula03, fragkoudi17}. At initial times, the bar of the intermediate disc is slightly stronger than the thin disc bar because of initial heating of the thin disc that makes it more stable than the intermediate disc, until 0.5~Gyr. The bar angular speed decreases with time by angular momentum transfer from the disc to the dark matter halo as seen on Fig~\ref{barst-fig} \citep[e.g.][]{debattista98, athanassoula02, dimatteo14}. The largest angular momentum transfer occurs from the thin disc to the halo (as in \citet[e.g.][]{fragkoudi17}). This slowing-down allows the bar to radially extend. It develops a buckling instability associated to decreases in the bar strength visible on Fig.~\ref{barst-fig} at $t=1.9$~Gyr and 3~Gyr. This buckling can be seen on the edge-on views of Fig~\ref{snaps-fig}. The bar speed is estimated by a Fourier method as in \citet[e.g.][]{halle15} and resonances radii (radii at which stars on nearly circular orbits resonate with the bar) are obtained by finding the radii where $\Omega(R)-\Omega_p$ and $\kappa(R)$ are commensurable, with $\Omega(R)$ and $\kappa(R)$ the angular speed and epicyclic frequency (respectively) of particles on almost circular orbits at $R$. Fig.~\ref{four-fig} shows the determination of the radii of the inner Lindblad resonance (ILR hereafter) where $\Omega(R)-\Omega_p=\dfrac{\kappa}{2}$, corotation where $\Omega(R)-\Omega_p=0$ and of the outer Lindblad resonance (OLR hereafter) $\Omega-\Omega_p=-\dfrac{\kappa}{2}$ at different times, and their time evolution.

\section{Radial migration}
\label{section-rad-migr}
\subsection{Blurring and churning}
\begin{figure*}
\centering
\resizebox{\hsize}{!}{
\includegraphics[width=13cm]{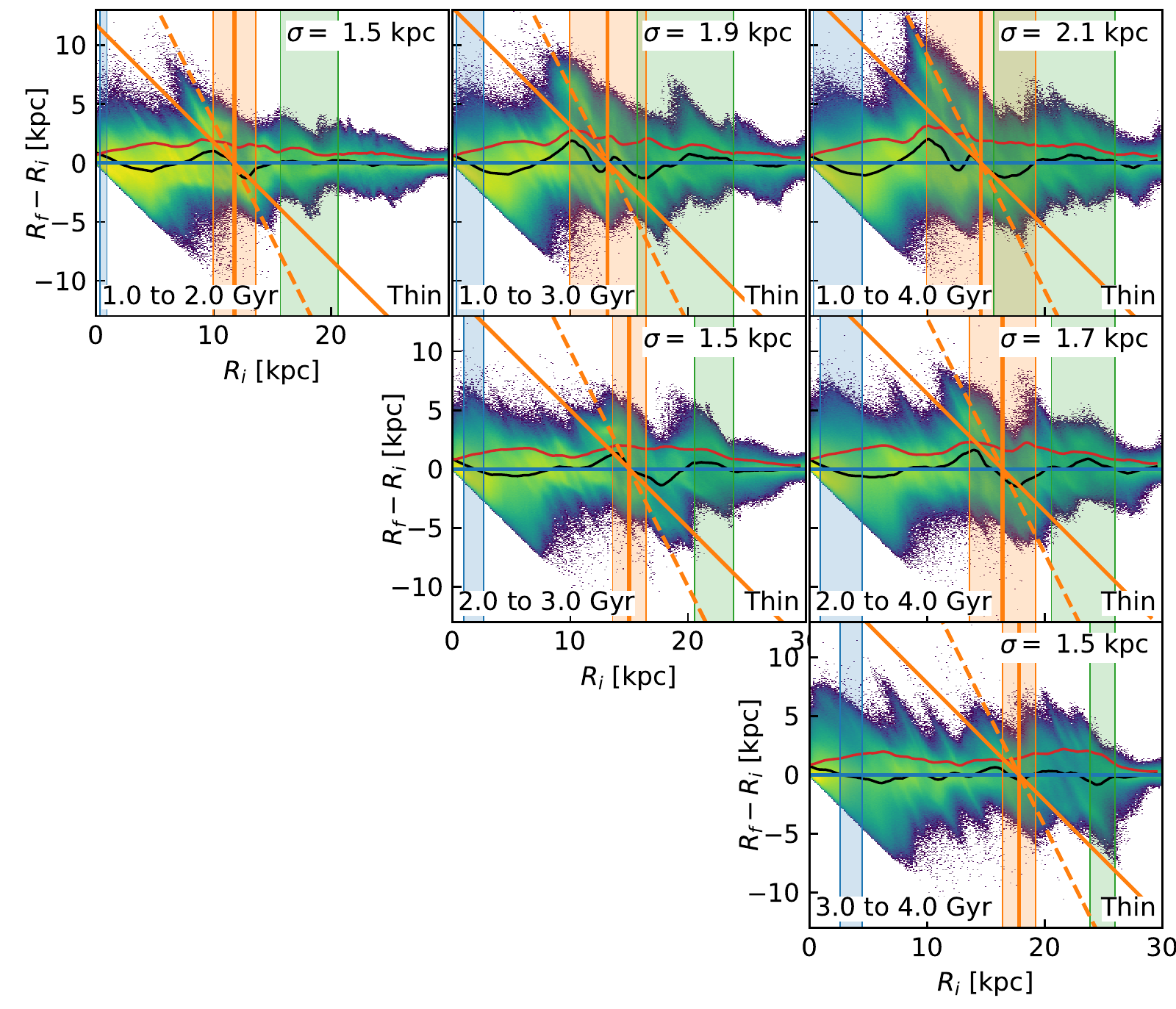} 
\includegraphics[width=13cm]{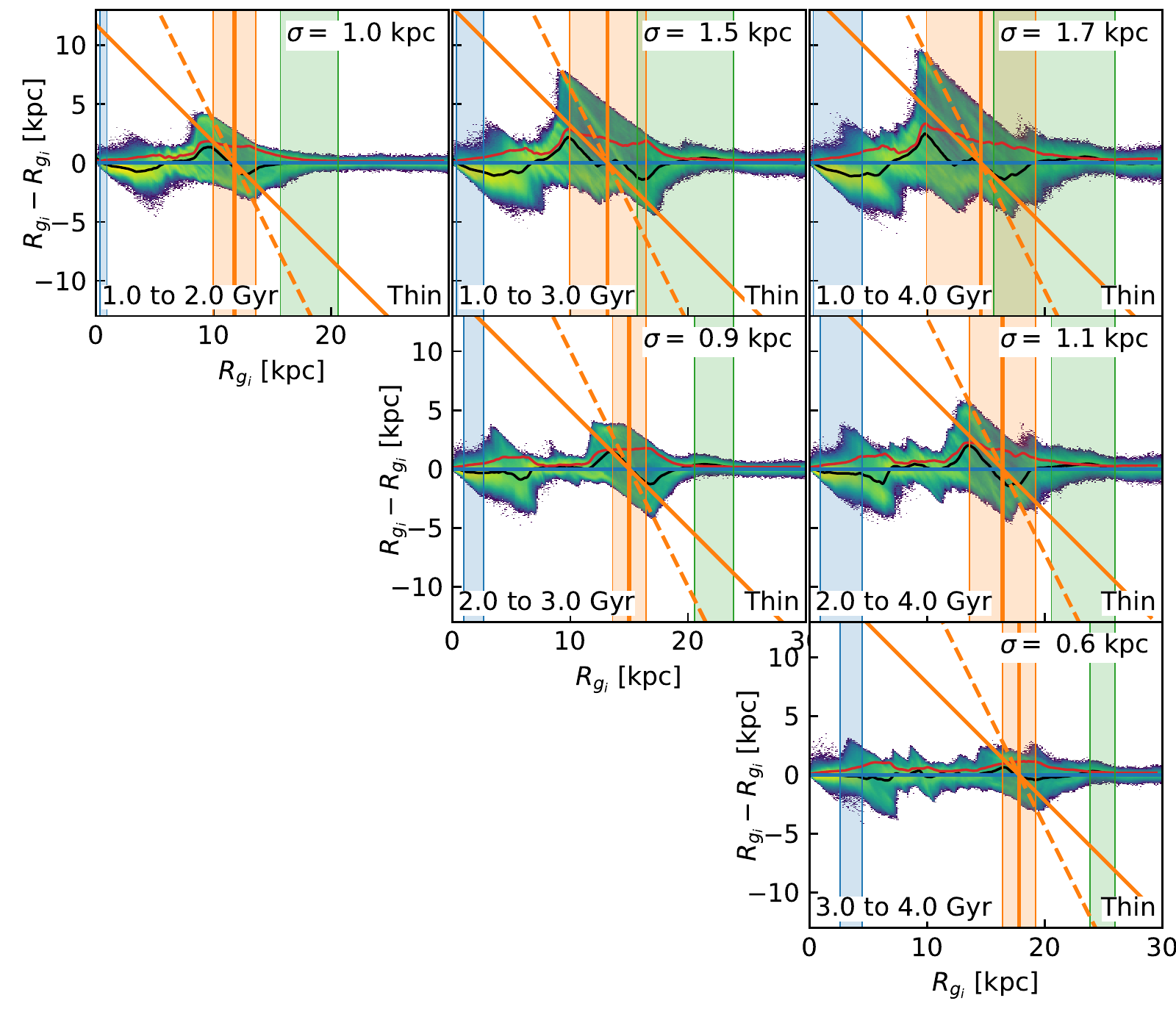} 
}
\caption{Left: Distributions of the change in radius $R_f-R_i$ as a function of $R_i$ for different time intervals in the thin disc. Right: Distributions of the change in guiding radius $R_{g_f}-R_{g_i}$ as a function of $R_{g_i}$ for different time intervals in the thin disc. On each panel, shaded regions show the values spanned by the resonances radii (blue for ILR, orange for corotation and green for OLR) during the time interval, and the thick vertical orange line shows the average of the values of the corotation radius at the beginning and end of the time interval.}
\label{radmigrthin-fig}
\end{figure*}

Radial migration is often separated into blurring, due to epicyclic oscillations around a guiding radius with time, and churning that is the change of the guiding radius \citep{sellwood02,schoenrich09}. We compute average radii (named guiding radii hereafter) as in \citet{halle15}, by a local average of the radial oscillations of a stellar particle using the closest local radial minima and maxima. 

Fig.~\ref{radmigrthin-fig} shows the distributions of the changes in galactocentric radius and guiding radius as a function of the initial values, for different time intervals. The black and red curves are the average and the dispersion (respectively) as a function of initial radius or guiding radius, and the total rms value is indicated in each panel. During these time intervals, the resonance radii grow (see Fig.~\ref{four-fig}), and the corresponding encompassed radial ranges are shown in shaded areas on Fig.~\ref{radmigrthin-fig}. A mean corotation radius (average of the values of the corotation radius at the beginning and end of the time interval) is represented by a thicker vertical line, together with a line of slope $-1$ intersecting the $x$-axis at the mean corotation radius, allowing to select stars crossing the mean corotation outwards or inwards, and a line of slope $-2$ with the same $x$-axis intersection at the mean corotation radius, along which stars exchange their position with respect to the mean corotation radius (as in \citet{sellwood02}). Some diagonal features are visible in the distributions, they are located around some resonances of the bar such as corotation, with outwards migration of stars below the resonance radius and inwards migration of stars beyond the resonance radius. For each pair of panels corresponding to the same time interval, the global dispersion and maximum change is lower for guiding radius because the epicyclic oscillations around the guiding radii are removed. 

As in many other studies \citep[e.g.][]{sellwood02}, we see the main churning occurs around corotation, with other patterns located at the ILR or for example the ultra-harmonic resonance $\Omega-\Omega_p=\dfrac{\kappa}{4}$ at $R \simeq 10$~kpc from $t=2$ to $t=3$~Gyr (only the ILR, corotation and OLR radii are shown on the panels). As explained in \citet{halle15}, a slow-down of the bar, implying a larger and larger corotation radius, allows for a large part of the disc to be affected by migration around an outwards-shifting corotation. 

Comparison of panels of change in guiding radius during the same duration also show that the amplitude of churning at corotation depends on the strength of the bar as in \citet{halle15}: it is for example stronger from $t=1$ to $t=2$~Gyr than from $t=3$ to $t=4$~Gyr (global rms value of 1~kpc and 0.6~kpc, respectively, and churning signal at corotation of a lower amplitude), as the bar is weaker at late times (see Fig.~\ref{barst-fig}). 

The top right panel of Fig.~\ref{radmigrthin-fig} show the change in guiding radius of thin disc stars on a time interval of 3~Gyr (from $t=1$ to $t=2$ Gyr). It can be seen that stars with initial guiding radii close to the initial corotation radius are the most extreme outwards migrators (their guiding radius can increase by as much as almost $10$~kpc). These extreme migrators will be studied in more details in section ~\ref{section-extreme-migr}.

\subsection{Migration in thick components}
\label{subsection-migrinthick}

\begin{figure*}
\centering
\includegraphics[width=15cm]{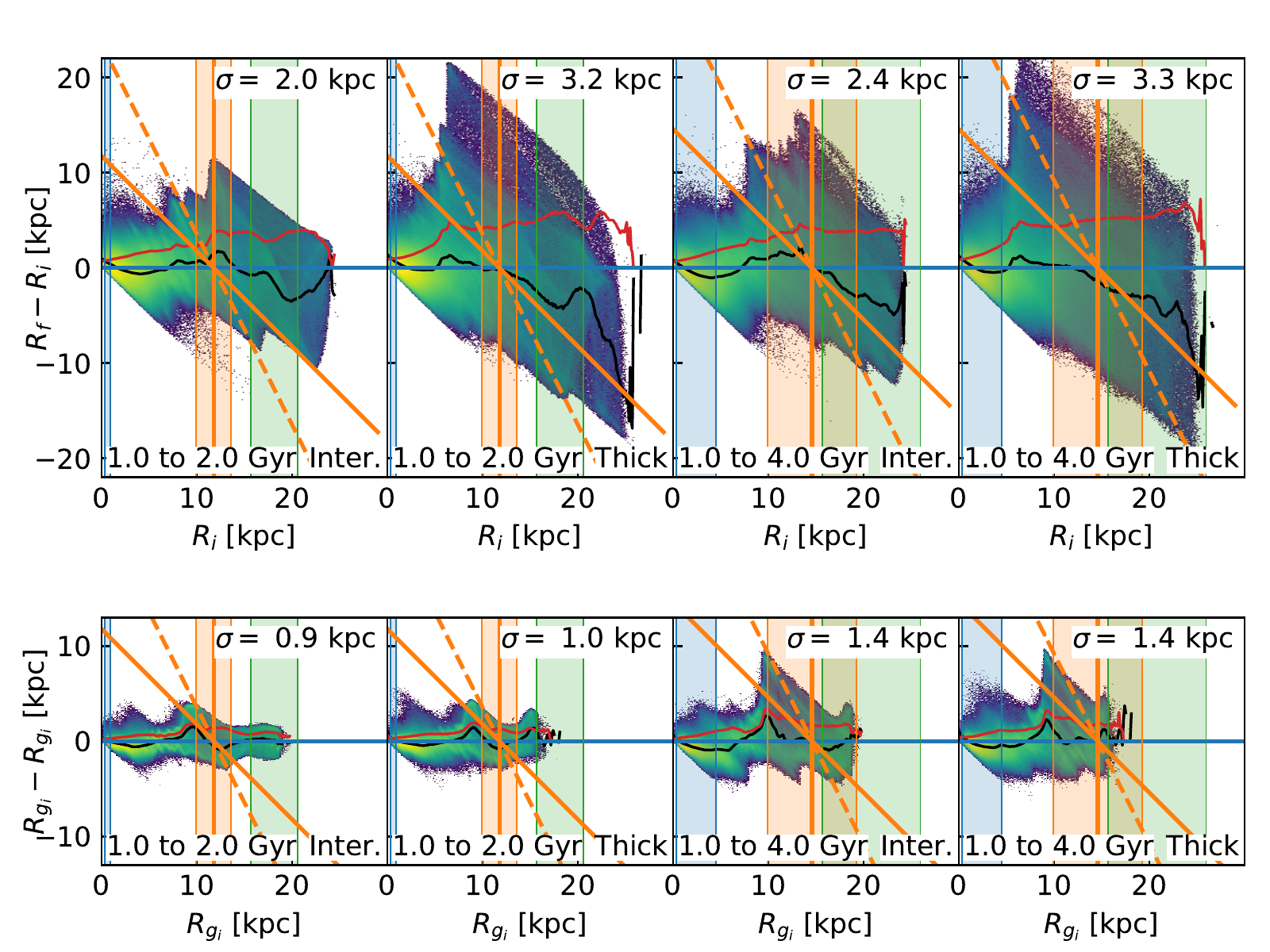}
\caption{First row: distributions of the change in radius $R_f-R_i$ as a function of $R_i$ from $t=1$~Gyr to $t=2$~Gyr (left panels) or from $t=1$~Gyr to $t=4$~Gyr (right panels) in the intermediate and thick discs. Second row: distributions of the change in guiding radius $R_{g_f}-R_{g_i}$ as a function of $R_{g_i}$ from $t=1$~Gyr to $t=2$~Gyr (left panels) or from $t=1$~Gyr to $t=4$~Gyr (right panels) in the intermediate and thick discs.}
\label{migrinterthick-fig}
\end{figure*}

Stars in the thick components have larger radial velocity dispersions than thin disc stars by construction (see Fig.~\ref{dispr-fig}) and thus larger radial excursions. The changes in galactocentric radius in a given time interval as from $t=1$ to $t=2$~Gyr represented on the two top-right panels of Fig~\ref{migrinterthick-fig} can thus be significantly larger than for the thin disc (top left panel of Fig.~\ref{radmigrthin-fig}). The largest migration in terms of galactocentric radius occurs at the OLR (for relatively few stars as this resonance is located near the outermost parts of the intermediate and thick discs). This migration signal at the OLR corresponds to stars migrating of as much as $\simeq$10~kpc in the intermediate disc, and as much as $\simeq$20~kpc in the thick disc. 

However, comparison of the bottom panels of Fig~\ref{migrinterthick-fig} to the corresponding plots on this time interval in Fig.\ref{radmigrthin-fig} shows that the changes in guiding radius are similar, both in their amplitude and RMS value (global ones or as a function of initial guiding radius (red curves)). This is consistent with results of \citep{solway12} that found similar changes in vertical angular momentum for thin and thick discs of simulated galactic discs with a bar. The relatively weak churning at OLR implies that the large migration in terms of galactocentric radius is due to the shape of orbits of particles at the OLR that allows them to have very large radial excursions. 

The higher radial velocity dispersion of the thick component is also associated with a larger asymmetric drift effect, implying that stars resonating with the bar are on average at a lower radius in the thick components than in the thin disc. The migration signal around corotation is thus slightly shifted to lower radii or guiding radii for thick components as can be seen for example on the bottom-right panel of Fig.~\ref{migrinterthick-fig} compared to the corresponding panel of Fig.~\ref{radmigrthin-fig}.

As for the thin disc, during in the 3~Gyrs time interval from $t=1$ to $t=4$~Gyr, some stars with an initial guiding radius close to the initial corotation radius can be churned outwards by almost $10$~kpc, which will also be discussed in more details in section ~\ref{section-extreme-migr}. At the exception of those stars, especially in the thick components, stars that are far from their initial guiding radius are more likely so due to blurring rather than churning effects.

\subsection{Comparison to a fixed potential}

\label{subsec-fixed-pot}

To compare the simulation to a case in which the bar has a constant strength and length, we take the gravitational potential of the simulation at a time $t_0$ and integrate orbits by simply rotating this fixed gravitational potential of an angle $\Omega_{P, t_0} (t-t_0)$ at each time $t$, with $\Omega_{P, t_0}$ the pattern speed at $t_0$, determined from the previous Fourier analysis of Section.~\ref{section-num-sim}. We use a Kick-Drift-Kick time integration with a timestep of $1$~Myr. Orbits are integrated from $t_0=1$~Gyr to $t=4$~Gyr and we compute their radius and guiding radius as a function of time as for the simulation particles. 

\begin{figure*}
\centering
\begin{tabular}{c}
\includegraphics[width=13cm]{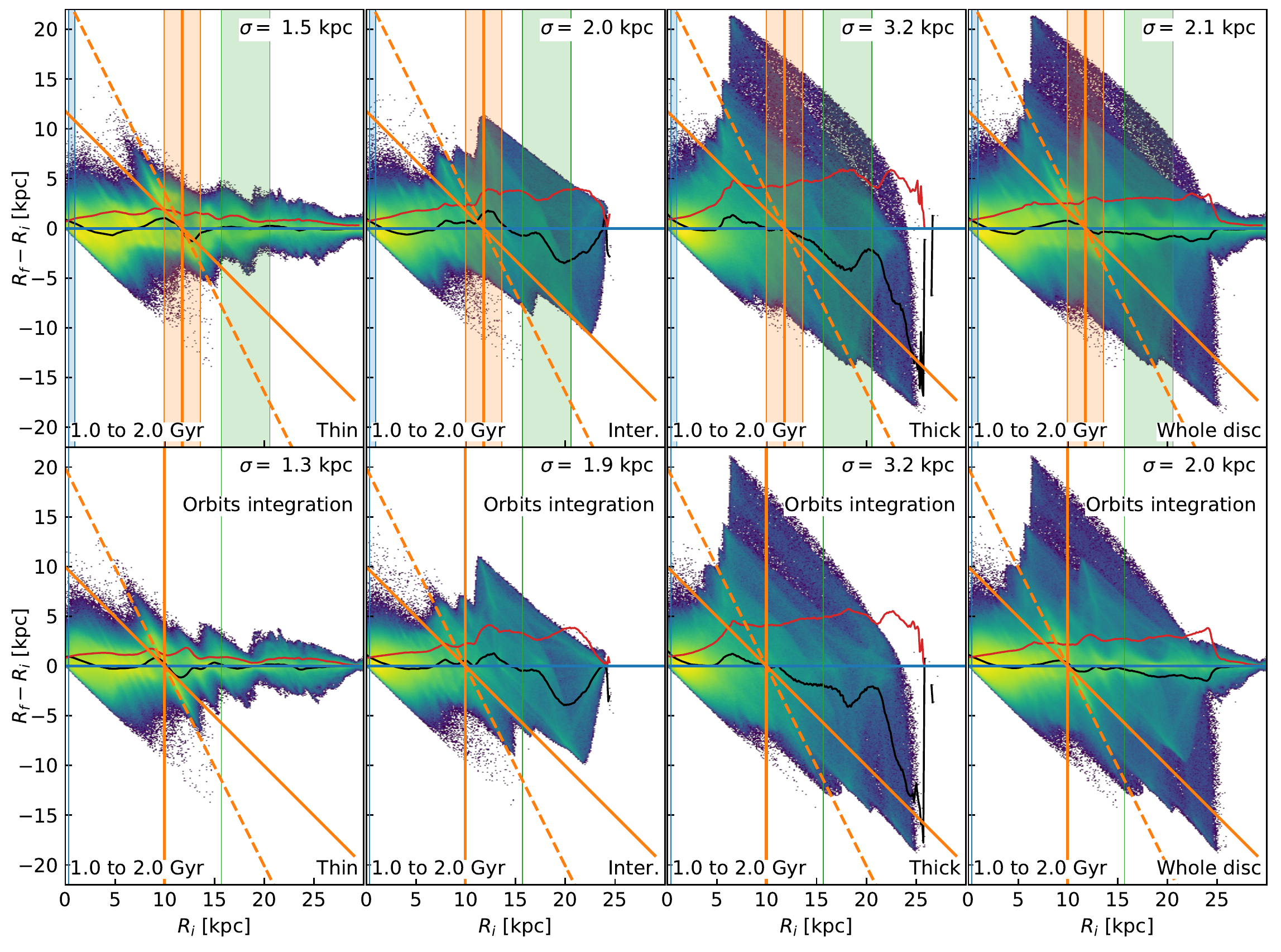} \\
\includegraphics[width=13cm]{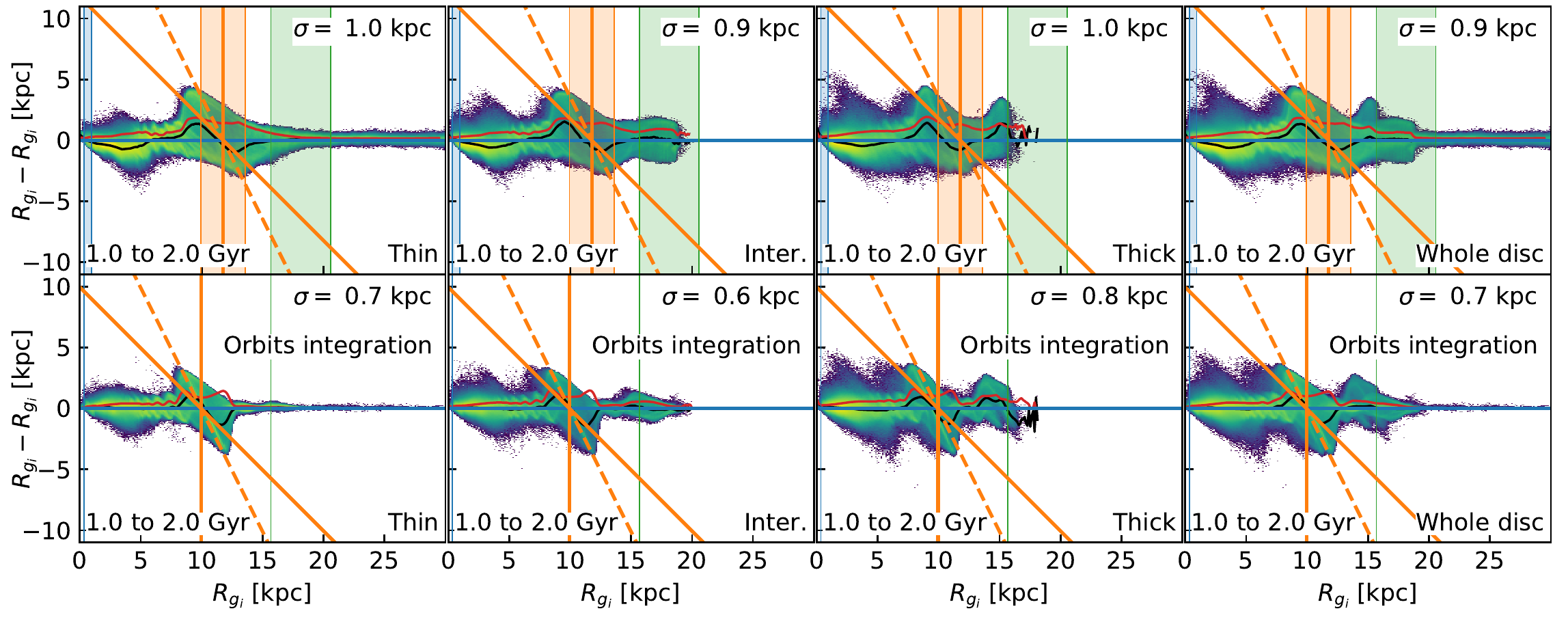} 
\end{tabular}
\caption{First and second rows: Distributions of the change in radius $R_f-R_i$ as a function of $R_i$ in the simulation (first row) and for the integration of orbits (second row) from $t=1$~Gyr to $t=2$~Gyr . Third and fourth rows: Distributions of the change in guiding radius $R_{g_f}-R_{g_i}$ as a function of $R_{g_i}$ in the simulation (third row) and for the orbits integration (fourth row) from $t=1$~Gyr to $t=2$~Gyr.}
\label{simorbs1to2-fig}
\end{figure*}

Fig.~\ref{simorbs1to2-fig} shows a comparison of radial migration in the simulation and for the orbits integration for the different disc components and the whole disc from $t=1$ to $t=2$~Gyr. During this time-interval, the corotation radius grows by a few kpcs (from 10~kpc to 13~kpc), which broadens the churning signal at corotation in the simulations in comparison to the orbits integration. More stars are churned in the shifting corotation region, which make the global rms values of churning larger in all components. The amplitude of churning are only mildly larger in the simulation than for orbits integration. 

\begin{figure*}
\centering
\begin{tabular}{c}
\includegraphics[width=13cm]{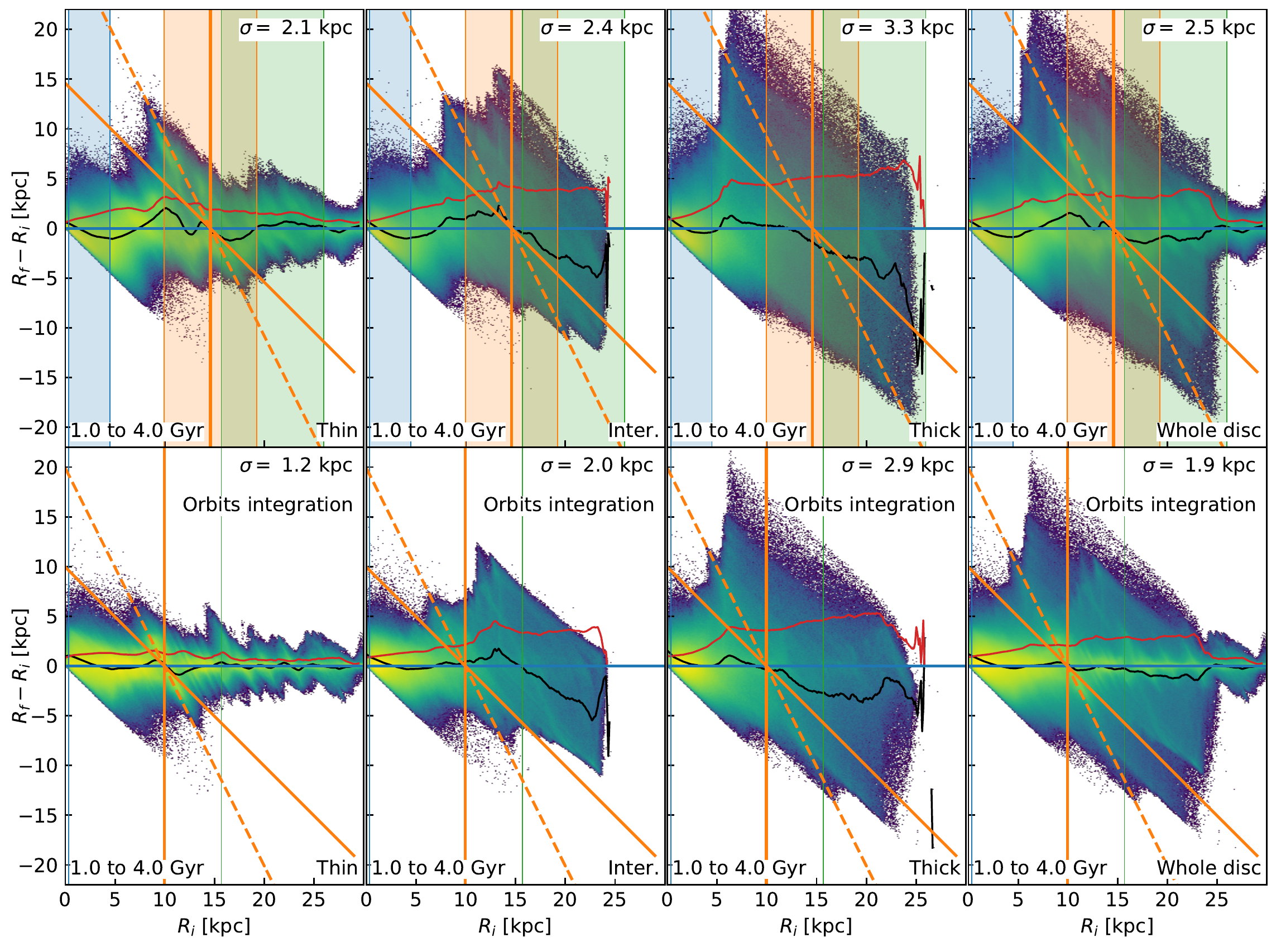} \\
\includegraphics[width=13cm]{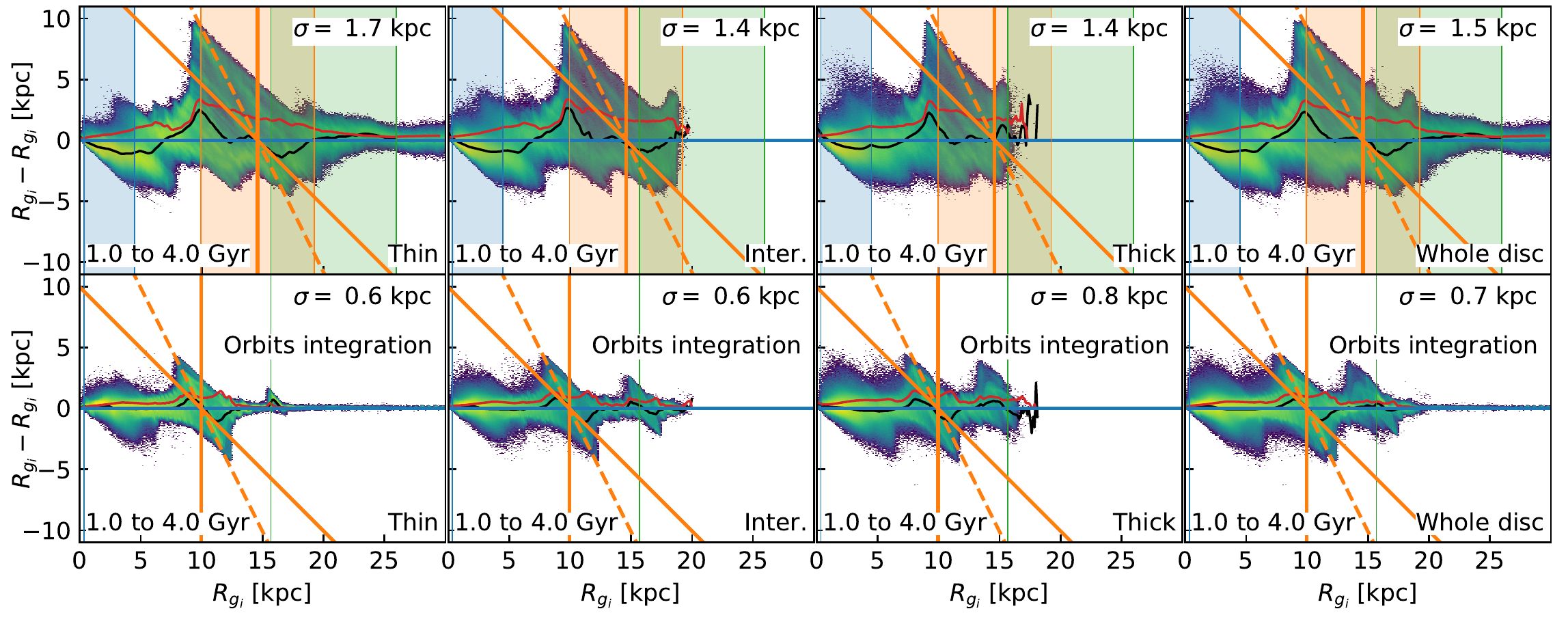} 
\end{tabular}
\caption{First and second rows: Distributions of the change in radius $R_f-R_i$ as a function of $R_i$ in the simulation (first row) and for the integration of orbits (second row) from $t=1$~Gyr to $t=4$~Gyr . Third and fourth row: Distributions of the change in guiding radius $R_{g_f}-R_{g_i}$ as a function of $R_{g_i}$ in the simulation (third row) and for the orbits integration (fourth row) from $t=1$~Gyr to $t=4$~Gyr.}
\label{simorbs1to4-fig}
\end{figure*}

Differences between the simulation and the integrated orbits in the fixed potential are more visible on a longer time-interval. Fig.~\ref{simorbs1to4-fig} shows a comparison between the simulation and orbits integration for a time interval from $t=1$ to $t=4$~Gyr. Unlike for the previous time-interval of 1~Gyr, a significant difference can be seen for both changes in galactocentric radii and guiding radii around corotation: stars can be churned outwards by almost twice as much a distance in the simulation compared to the orbits integration, for all disc components. 

In the case of a fixed pattern rotating at a fixed speed, stars can remain trapped at corotation in a libration movement around the local maxima of the effective potential $\phi_{\rm eff}=\phi_g-1/2 \Omega_p^2 R^2$, where $\phi_g$ is the gravitational potential and $\Omega_p$ the pattern speed (see Section3.3 of \citet{binneytrem08} and \citet{sellwood02}). The time evolution of the galactocentric radius of stars trapped at corotation consists of oscillations around a more slowly oscillating guiding radius. Here, the changes in guiding radius in the orbits integration are fairly similar in both considered time-intervals, both in amplitude and rms values because of these churning oscillations of trapped stars (the first time-interval is already similar to or larger than the average half-period of oscillations of guiding radius around corotation). For an evolving pattern-speed, the situation is more complicated than periodic churning as stars can remain trapped, but also be liberated along with the shifting of the resonance location, and new stars can become trapped. In the new section we focus on the extreme migrators, stars churned outwards the most.  

\section{Extreme migrators}

\label{section-extreme-migr}

\subsection{Stars at corotation}

\label{subsec-crparts}

\begin{figure*}
\centering
\resizebox{\hsize}{!}{\includegraphics[width=13cm]{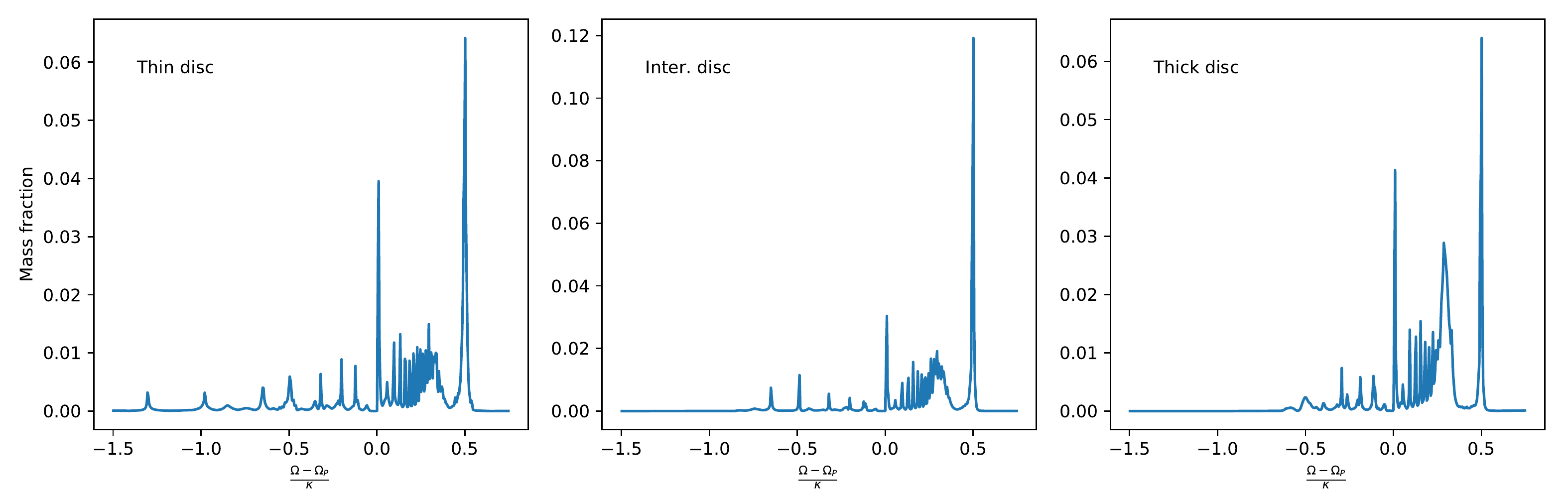}} 
\caption{Ratio $\dfrac{\Omega-\Omega_p}{\kappa}$ with $\Omega_p$ the bar speed at $t=1$~Gyr}
\label{raps-fig}
\end{figure*}

\begin{figure*}
\centering
\resizebox{\hsize}{!}{
\begin{tabular}{ccc}
\includegraphics{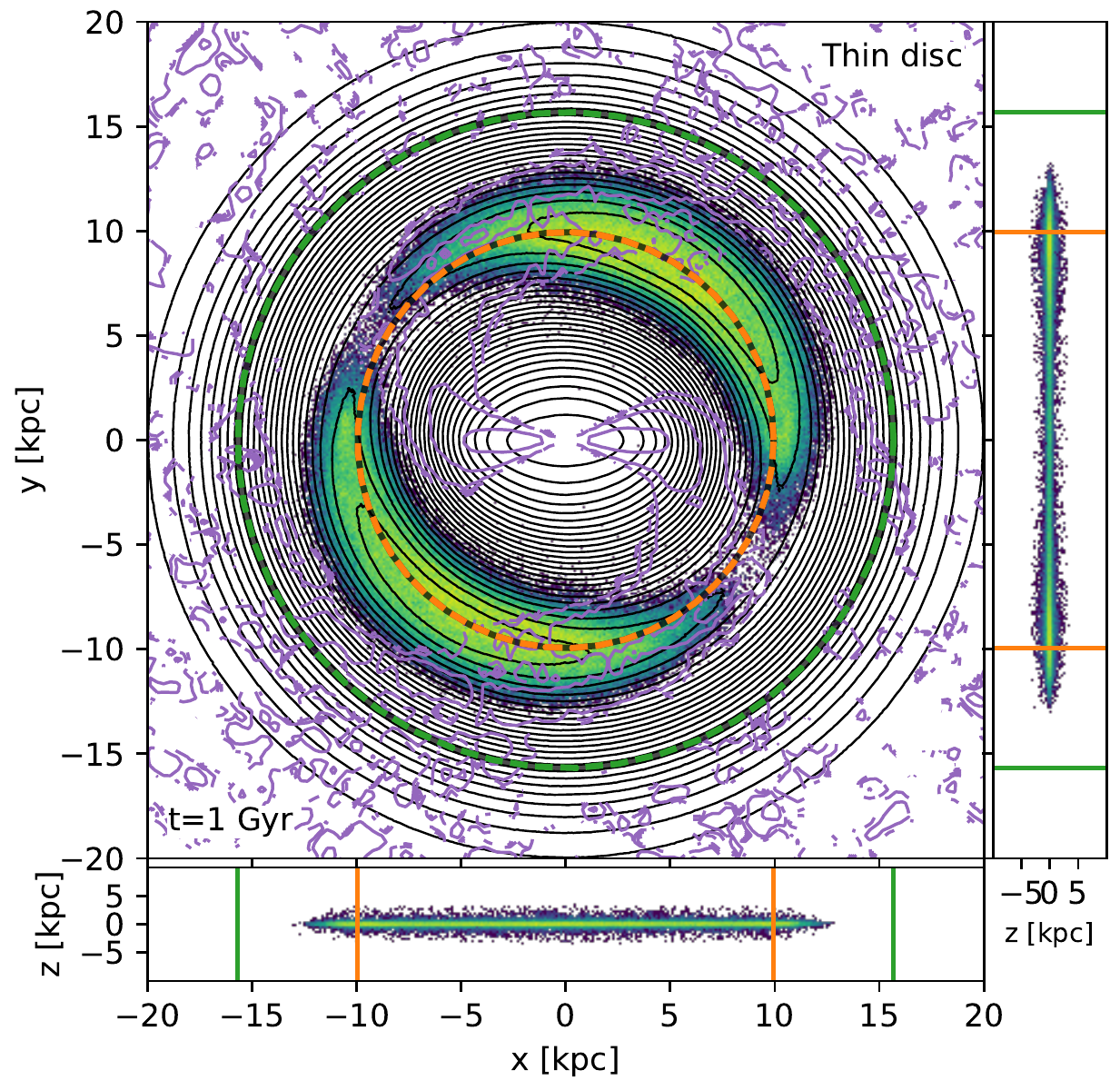} & \includegraphics{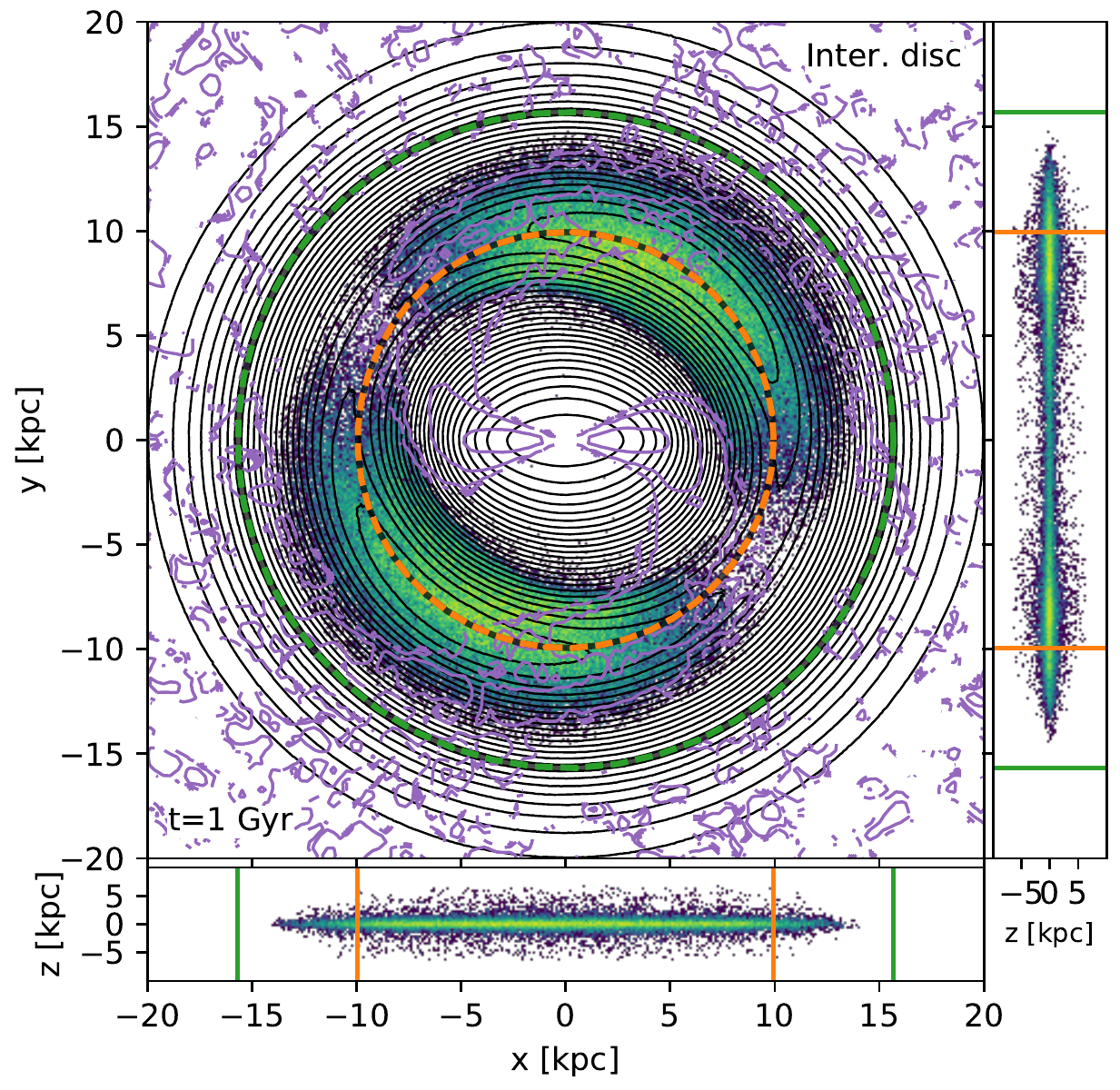} 
& \includegraphics{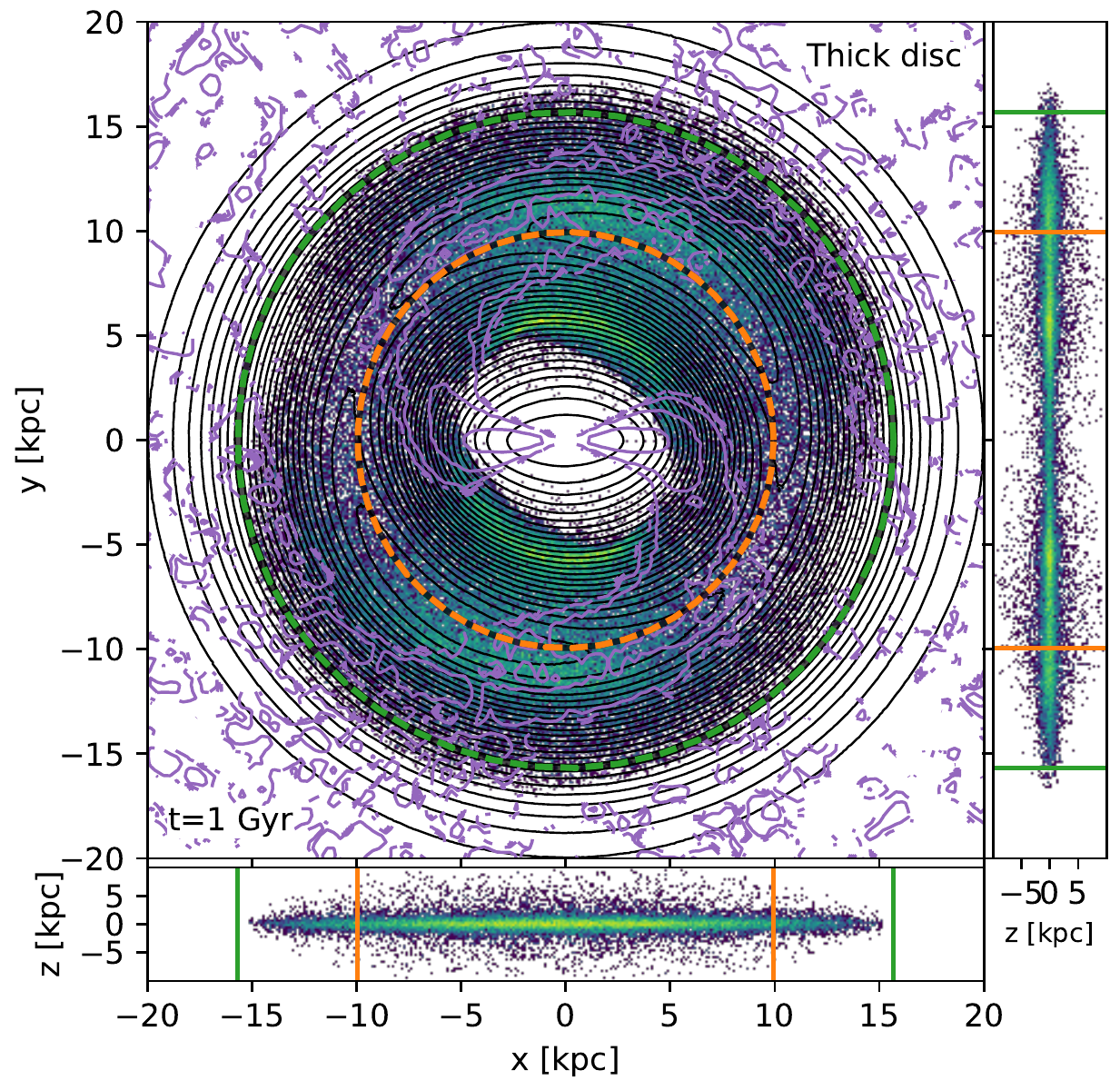} 
\end{tabular}
}
\caption{Colours: density map of stars corotating with the bar. Black contours: effective potential in a frame rotating at the bar speed. Purple contours: positive azimuthal overdensity in radial bins.}
\label{loccr-fig}
\end{figure*}

We first look for stellar particles corotating with the bar at $t=1$~Gyr, so as to study their migration. Determining which stars resonate with a non-axisymmetric pattern is possible by extracting their individual angular $\Omega$ and radial $\kappa$ frequencies from a Fourier analysis of their orbits (see \citet{binneysperg82,athanassoula02,ceverino07}). We analyse orbits integrated with the simulation gravitational potential of $t=1$~Gyr rotating at the $t=1$~Gyr bar speed as described in~\ref{subsec-fixed-pot}, to obtain stars resonating with the bar at $t=1$~Gyr. We compute the radial frequency $\kappa$ from a Fourier transform of the time evolution of the radius, removing the (usually) low frequency of the angular momentum oscillations, and estimate the angular frequency $\Omega$ by a Fourier transform of the Cartesian $x$ or $y$ coordinates divided by the radius. Fig.~\ref{raps-fig} shows the distribution of the ratio $\frac{\Omega-\Omega_p}{\kappa}$, with $\Omega_p$ the bar speed at $t=1$~Gyr. Several peaks can be seen, the most prominent ones being at the ILR ($\frac{\Omega-\Omega_p}{\kappa}=0.5$) and corotation ($\frac{\Omega-\Omega_p}{\kappa}=0$) in all disc components. There is also a complex structure of peaks between corotation and ILR concerning a large fraction of the stars, especially in the thickest component for which a high and broad peak is visible. This is likely due to the beginning of the buckling instability and is similar to the patterns obtained in \citet{martinezvalp06}. The OLR peak ($\frac{\Omega-\Omega_p}{\kappa}=-0.5$) is also visible in all three components. The surface density declines approximately exponentially with radius and the bar is strong, hence the larger fraction of stars at the ILR peak for all three components.

Fig.~\ref{loccr-fig} shows in colours the density of particles found at corotation by the Fourier analysis, i. e. the particles of the $\dfrac{\Omega-\Omega_p}{\kappa}=0$ peak (making up $7.6 \%$ of the thin disc mass, $6 \%$ of the intermediate disc mass and $7.8 \%$ of the thick disc mass). The black contours are some isocontours of the effective potential in the frame rotating at the bar speed. The purple contours are positive isocontours of the overdensity of the whole disc as in the left column of Fig.~\ref{snaps-fig}. The bar is aligned with the $x$~axis. The local maxima of the effective potential (encompassed by closed curves in the top right and bottom left) are not on a line orthogonal to the bar as Lagrange points L4 and L5 of a simple barred potential, because the gravitational potential is gradually tilted by the spiral arms contribution as radius increases. Thin disc stars at corotation are localised around those local maxima. The distribution of stars of thicker components is wider in radius because of the higher eccentricities of their orbits. Figs.~\ref{locilrr-fig} and \ref{locolr-fig} show the density maps of stars at the ILR and at the OLR (respectively).

\subsection{Orbits of stars at corotation for a fixed potential rotating at a constant speed}

\begin{figure}[h!]
\centering
\resizebox{\hsize}{!}{\includegraphics{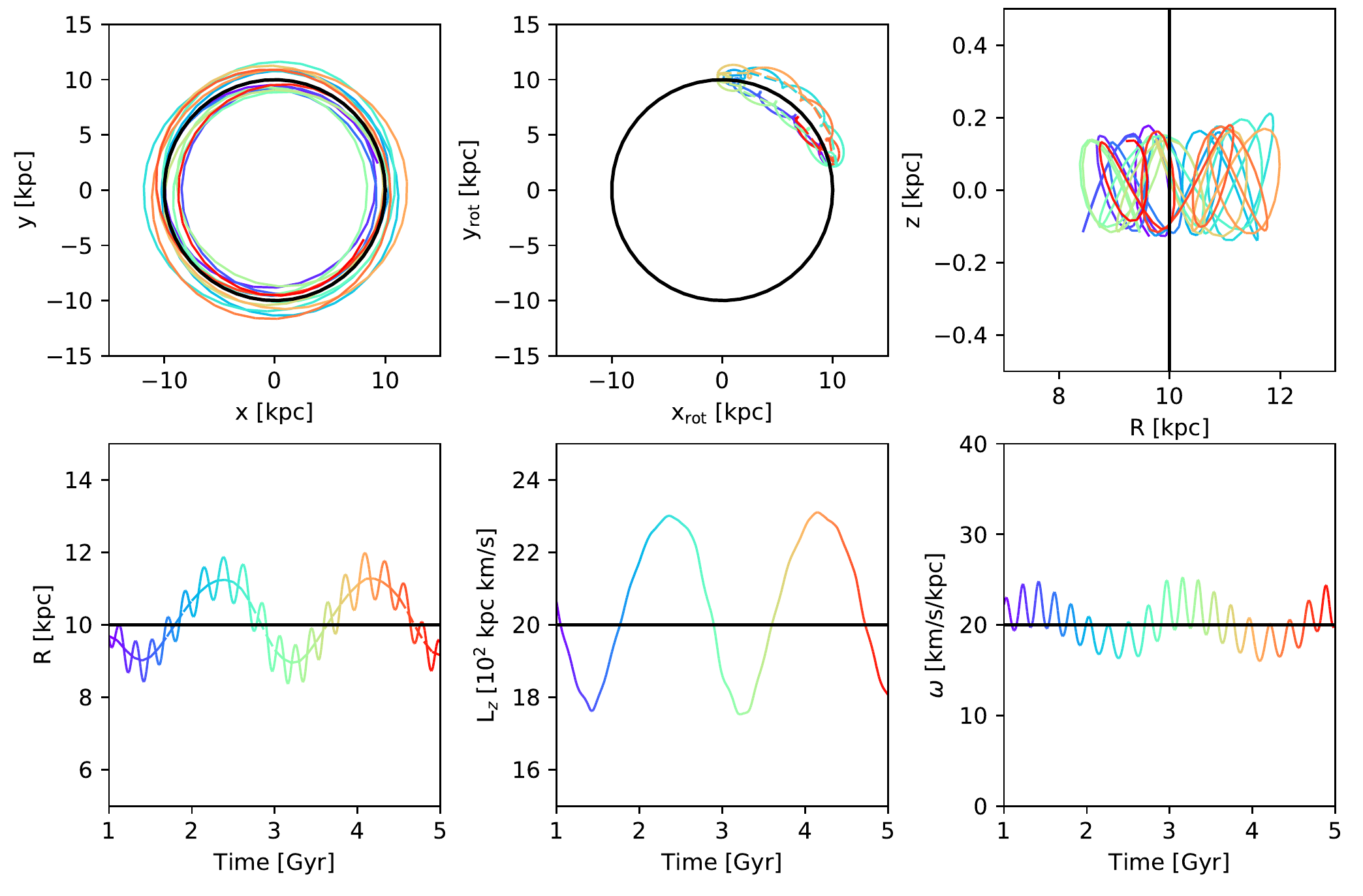}}
\caption{Thin disc stellar particle evolution in the orbits integration case. $x_{\mathrm{rot}}$ and $y_{\mathrm{rot}}$ are coordinates in a frame rotating around the $z$-axis at the pattern speed $\Omega_p$. Time is colour-coded with the same scale for all panels. Black curves: corotation radius, corotation $L_z$ (bottom-middle panel), and $\Omega_p$ (bottom-right panel).}
\label{thinorb-fig}
\end{figure}

Fig.~\ref{thinorb-fig} shows an example orbit of thin-disc stellar particles corotating with the bar: the orbit is represented by its $x$ and $y$ coordinates, and by $x_{\mathrm{rot}}$ and $y_{\mathrm{rot}}$, its coordinates in a frame rotating around the $z$-axis at the pattern speed $\Omega_p$. In both latter plots, the bar is parallel to the $x-$axis. In the rotating frame, the libration movement is clearly visible (the amplitude of the azimuthal excursions in the rotating frame vary from star to star). The maximum value of the churning amplitude is expected to increase with the strength of the non-axisymmetric patterns \citep{binneytrem08, sellwood02}.The stellar particle rotates either faster than the bar at a lower radius or slower at a larger radius (and on average at the bar speed). Its guiding radius is plotted with a dashed line, exhibiting the churning oscillations happening at corotation. The evolution of the radius as a function of time (bottom left panel) also shows epicyclic oscillations modulated by a lower frequency guiding radius oscillation around the corotation radius. 

\begin{figure}[h!]
\centering
\resizebox{\hsize}{!}{\includegraphics{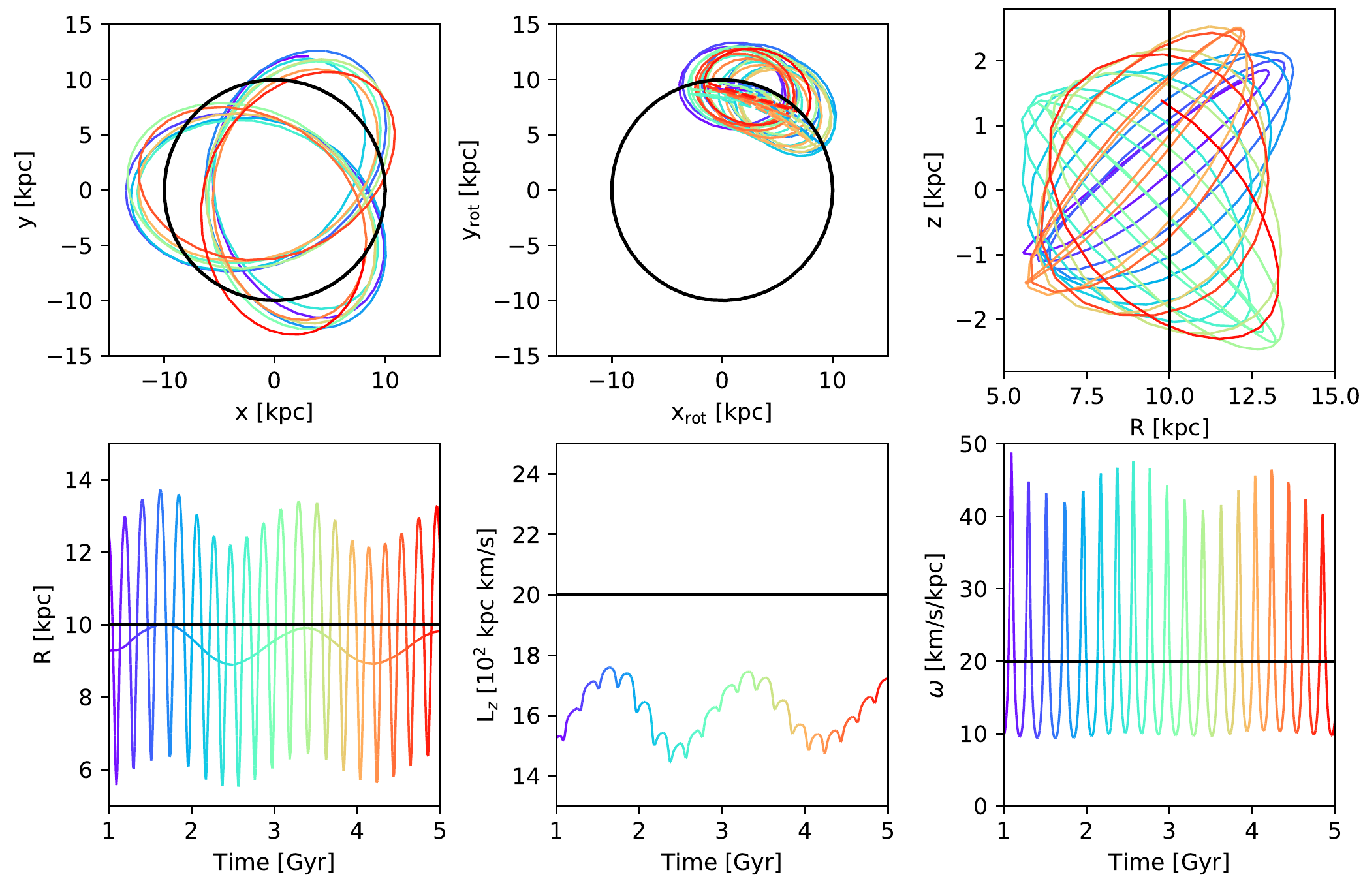}}
\caption{Same as Fig.~\ref{thinorb-fig} for a thick disc stellar particle.}
\label{thickorb-fig}
\end{figure}

\begin{figure*}
\centering
\begin{tabular}{c}
\includegraphics[width=15cm]{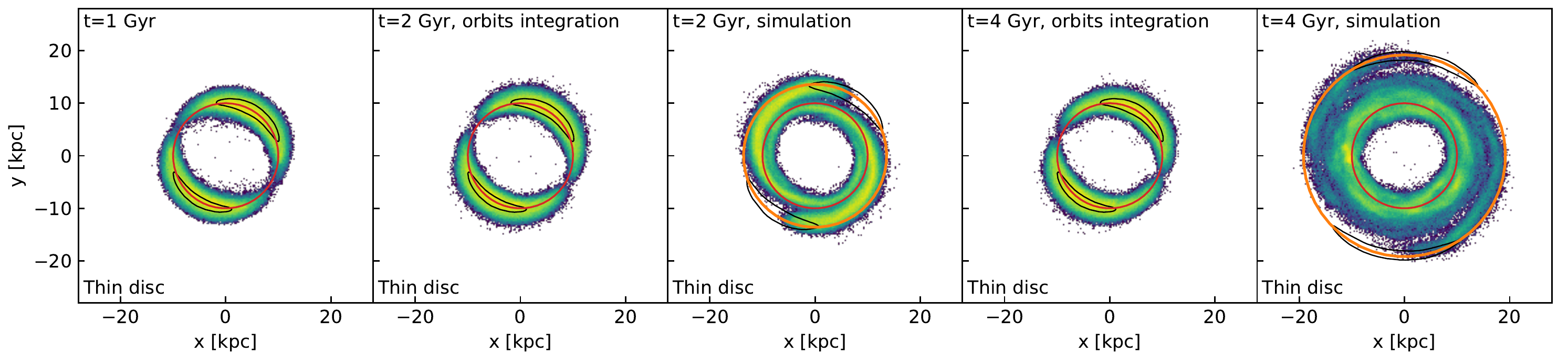} \\
\includegraphics[width=15cm]{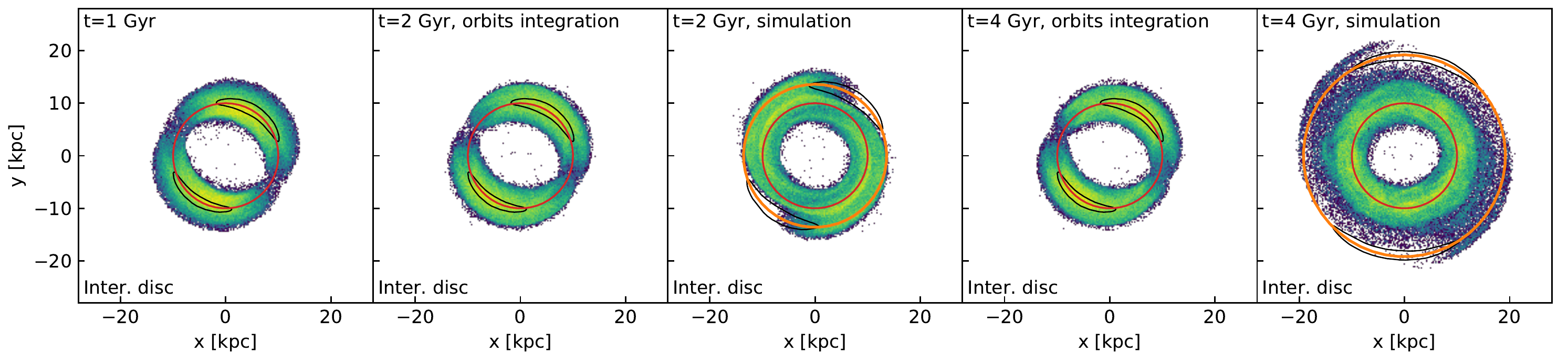} \\
\includegraphics[width=15cm]{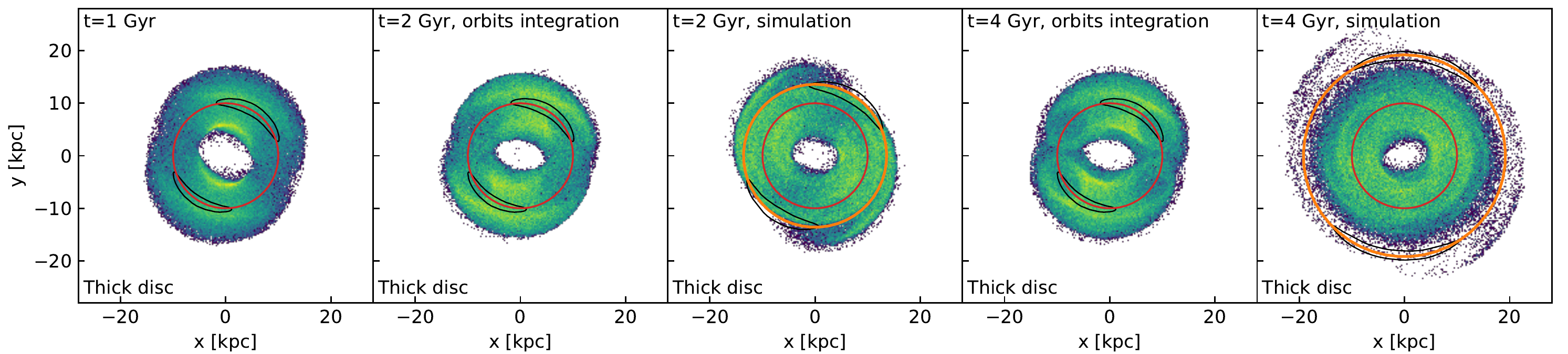}
\end{tabular}
\caption{Stars corotating with the bar at t=1~Gyr shown at  $t=1$~Gyr (initial time for the orbits integration) and at $t=2$~Gyr and $t=4$~Gyr in the orbits integration case and simulation. Maps are each rotated so that the bar is parallel to the $x-$axis. Red circle: corotation radius at $t=1$~Gyr. Orange circle: corotation radius at $t=2$ or $t=4$~Gyr. Black contours: isocontours encompassing the local maxima of the effective potential with the pattern speed at $t=1$~Gyr for the orbits integration panels, or at $t=2$ or $t=4$~Gyr for the simulation panels.}
\label{partsatcr-fig}
\end{figure*}

Stars of thicker components can also have librating orbits as see on Fig.~\ref{thickorb-fig}. Their radial "epicycle" excursions (the orbits are far from close to circular orbits) have an amplitude similar to the global radial motion amplitude. Because of the asymmetric drift effect discussed in section~\ref{subsection-migrinthick}, corotating stars are on average at lower radii than thin disc corotating stars, and have thus an average lower guiding radius and vertical angular momentum. Their average position in the rotating frame can also be closer to orthogonal to the bar in the gravitational potential of this orbits integration, because of the decreasing spiral tilting of the potential as radius decreases.

\subsection{Migration of stars at corotation at an initial time}

We now compare the fate of corotating stars at $t=1$~Gyr in the orbits integration case (with a constant bar speed) to the simulation case with a decreasing bar speed. 

Fig.~\ref{partsatcr-fig} shows the distribution of these stars at $t=2$ and $t=4$~Gyr in the orbits integration and in the simulation. Density maps are each rotated so that the bar is parallel to the $x$-axis. In the orbits integration, the distribution remains almost the same: stars just evolve on their orbits filling the same spatial area. 

In the simulation, the distribution is however torn out into a spiralling shape, with stars reaching higher radii, similar to the corotation radius at the end of the time-interval for the ones migrating the most (or even a few kpcs higher in the thick disc), and some stars remaining close to the $1$~Gyr corotation radius. A fraction of the stars thus remains trapped at corotation during the time-interval and migrates outwards, while the rest of the stars remain close to their initial radius. We determine which stars corotate with the bar at $t=2$~Gyr and $t=4$~Gyr by the same method as in~\ref{subsec-fixed-pot} and~\ref{subsec-crparts}, and find that the fraction of stars at corotation at $t=1$~Gyr remaining trapped at $t=2$ or $t=4$~Gyr decreases with the thickness of the disc component. $19 \%$ of the thin disc stars corotating with the bar at $t=1$~Gyr still corotate with the bar at $t=2$~Gyr, while this fraction is of $16 \%$ for the intermediate disc and $13 \%$ for the thick disc. Between $t=2$ and $t=4$~Gyr, some stars do not remain in corotation, only $4.8 \%$ of the thin disc stars corotating with the bar at $t=1$~Gyr still corotate with the bar at $t=2$~Gyr, while this fraction is of $3.9 \%$ for the intermediate disc and $1.2\%$ for the thick disc. 

\subsection{Extreme migrators}
\begin{figure}
\centering
\resizebox{\hsize}{!}{
\begin{tabular}{c}
\includegraphics{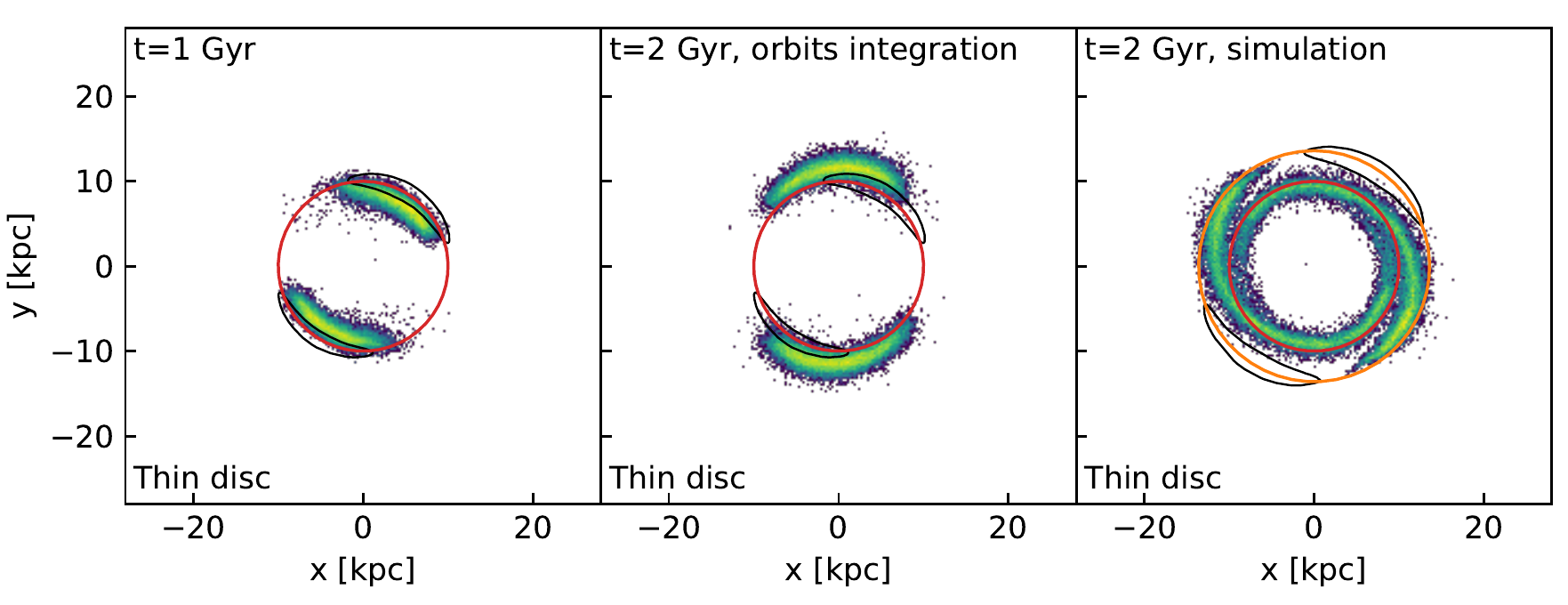} \\
\includegraphics{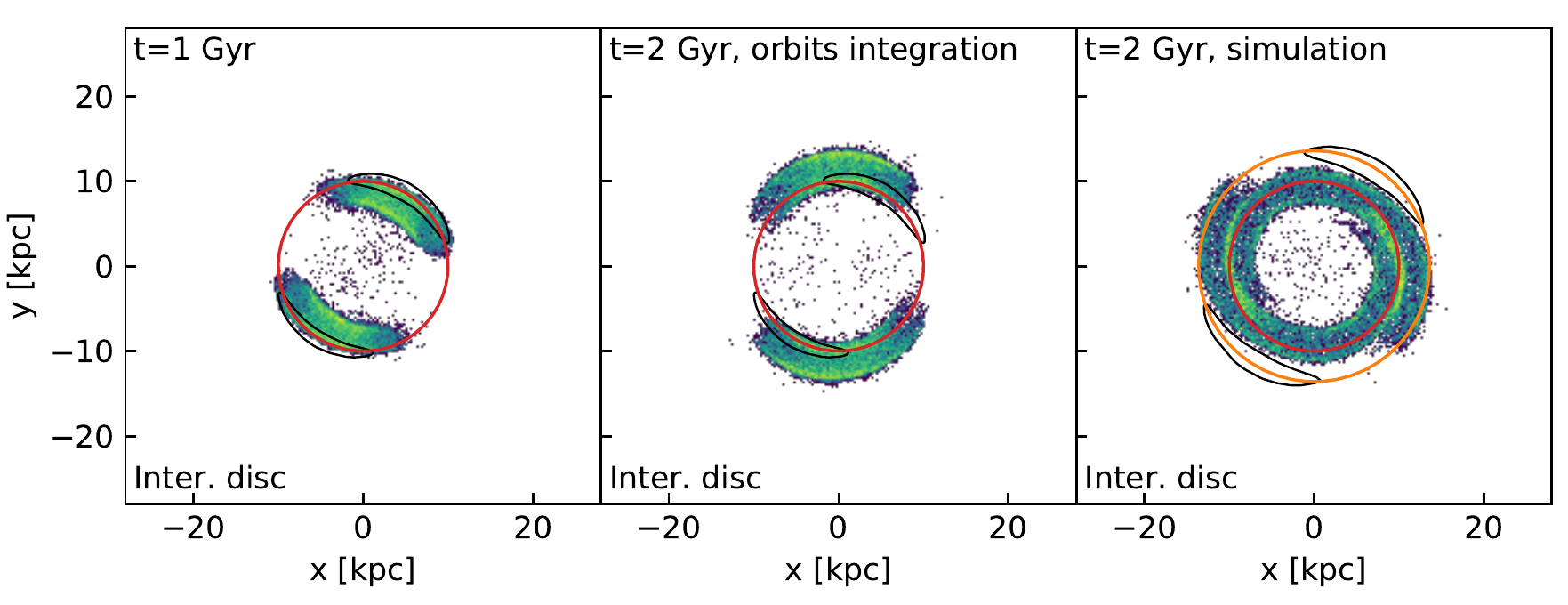} \\
\includegraphics{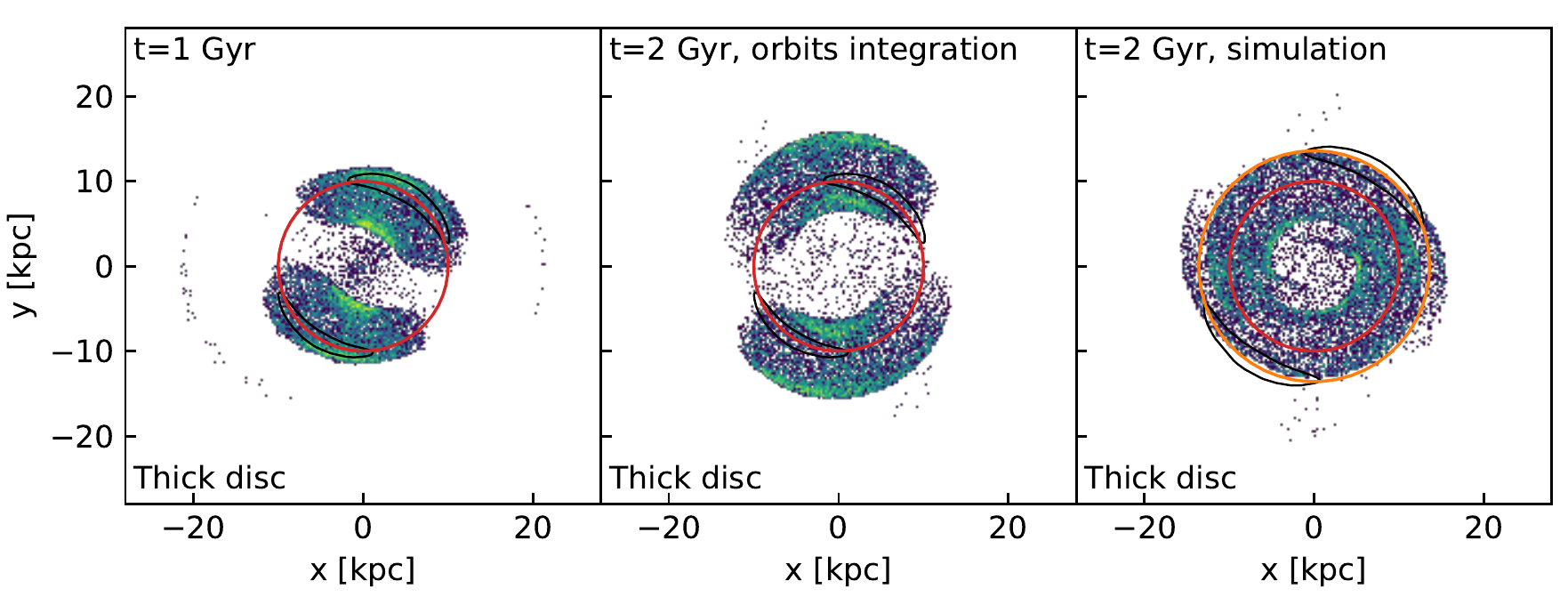} \\
\includegraphics{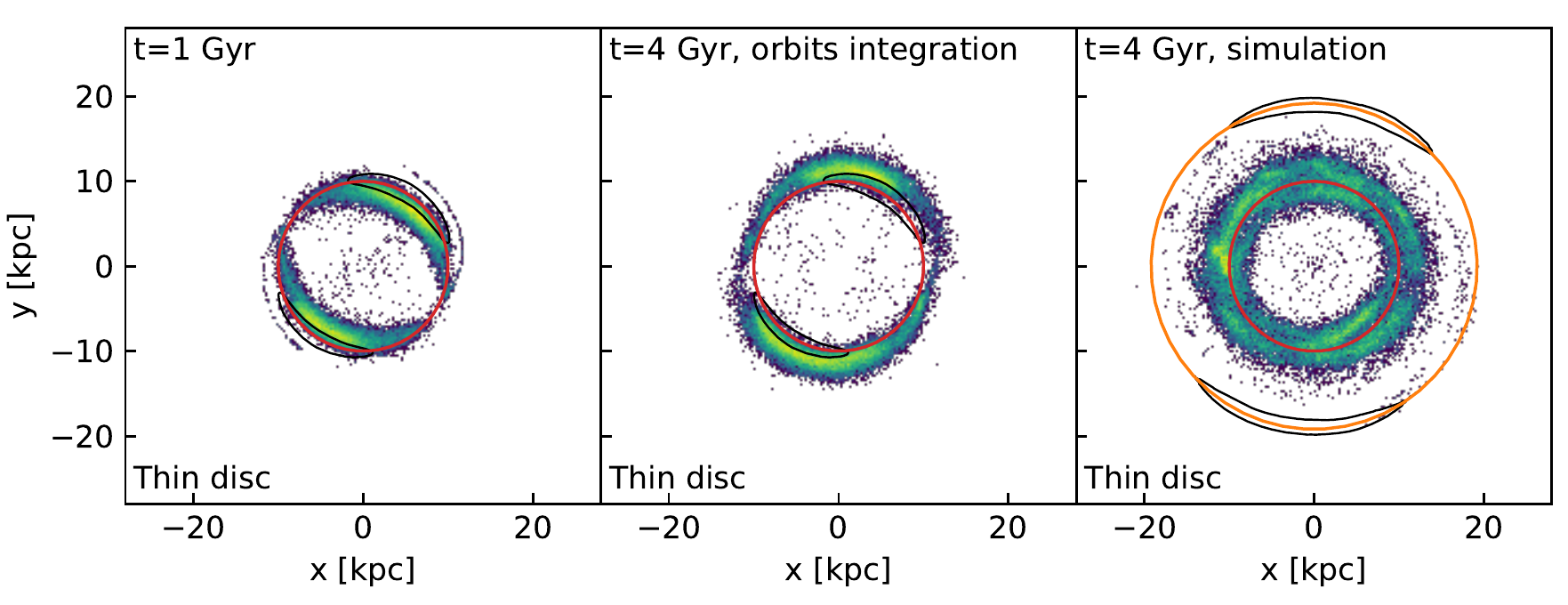} \\
\includegraphics{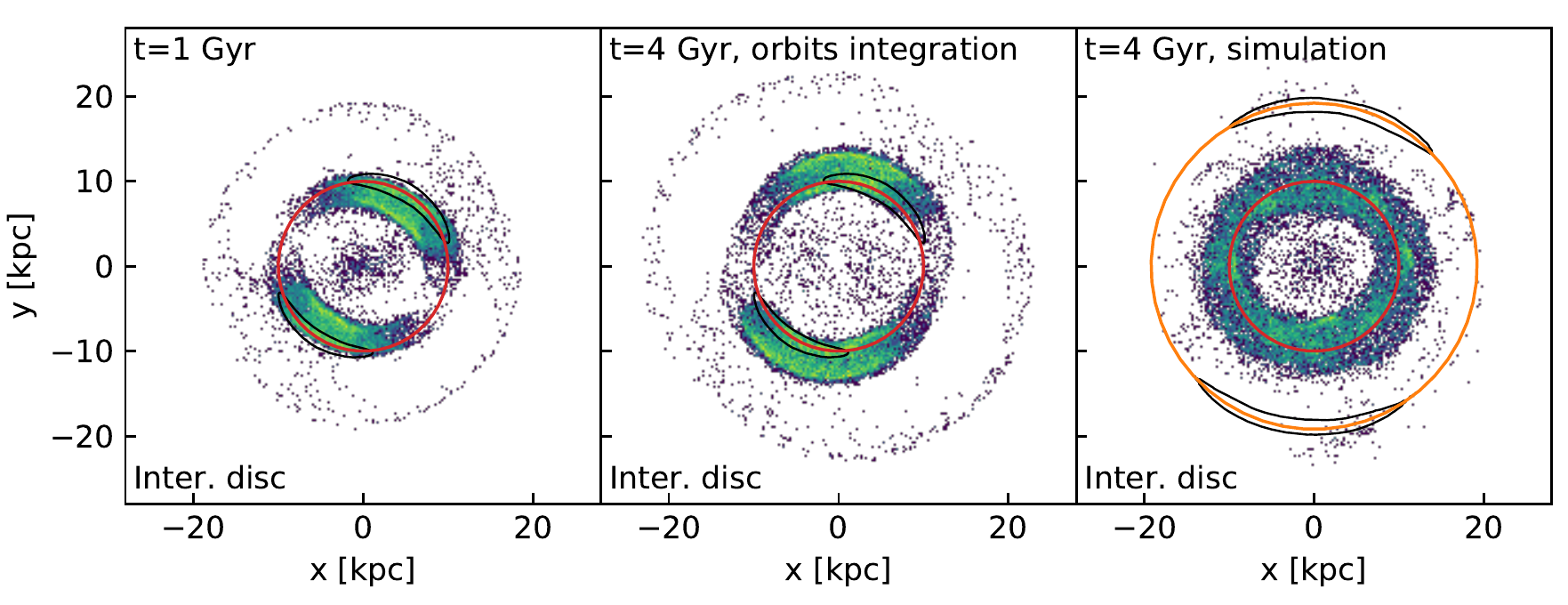} \\
\includegraphics{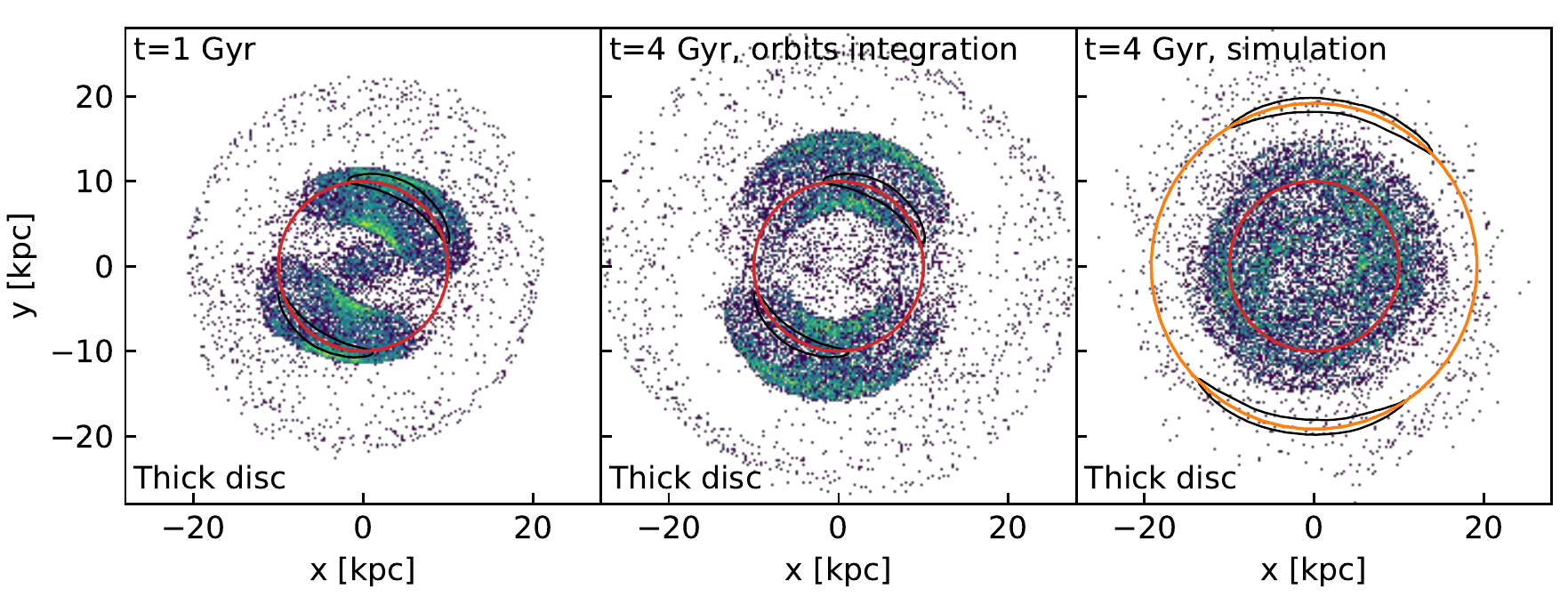} \\
\end{tabular}
}
\caption{Stars churned outwards the most in the orbits integration (top 1$\%$) from $t=1$ to $t=2$~Gyr (top three rows) or from $t=1$ to $t=4$~Gyr (bottom three rows). Left column: spatial distribution of these extreme migrators at $t=1$~Gyr. Middle and right columns: spatial distribution at $t=2$ or $t=4$~Gyr in the orbits integration case and in the simulation. Maps are each rotated so that the bar is parallel to the $x-$axis. Red circle: corotation radius at $t=1$~Gyr. Orange circle: corotation radius at $t=2$ or $t=4$~Gyr. Black contours: isocontours encompassing the local maxima of the effective potential with the pattern speed at $t=1$~Gyr for the orbits integration panels, or at $t=2$ or $t=4$~Gyr for the simulation panels.}
\label{extrmigrorbs-fig}
\end{figure}

\begin{figure}
\centering
\resizebox{\hsize}{!}{
\begin{tabular}{c}
\includegraphics{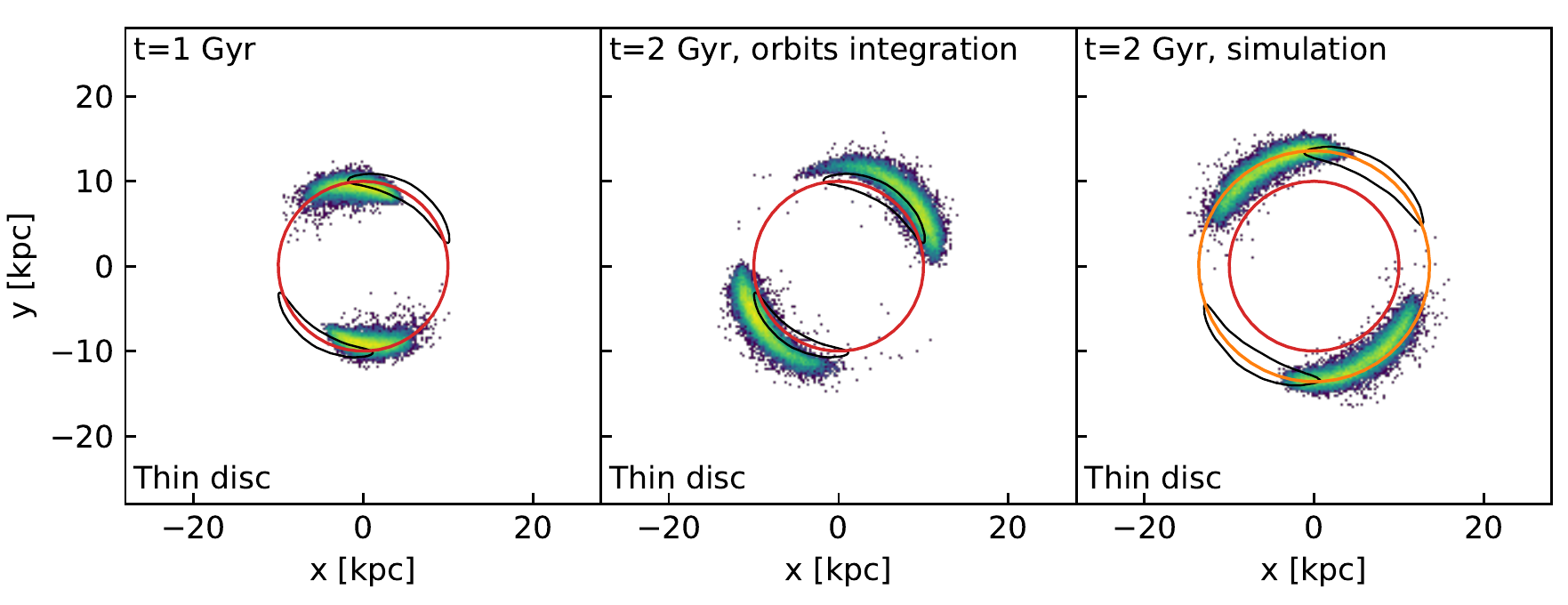} \\
\includegraphics{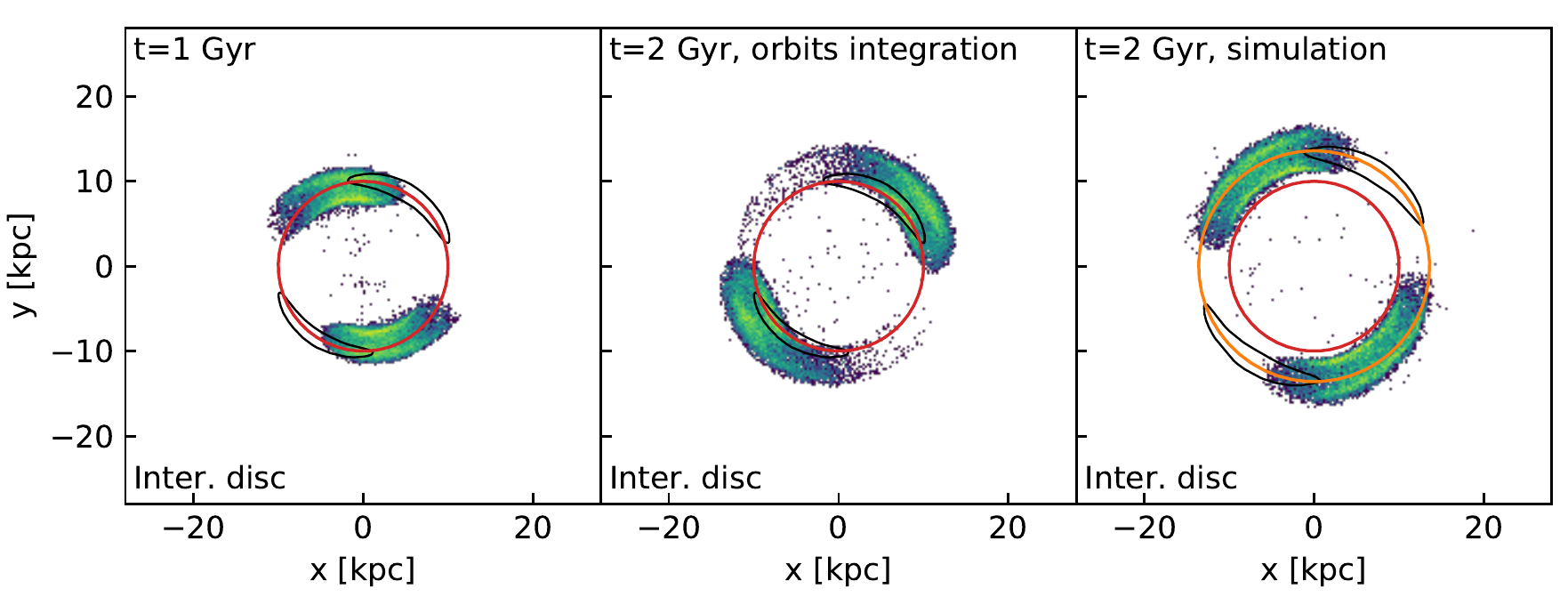} \\
\includegraphics{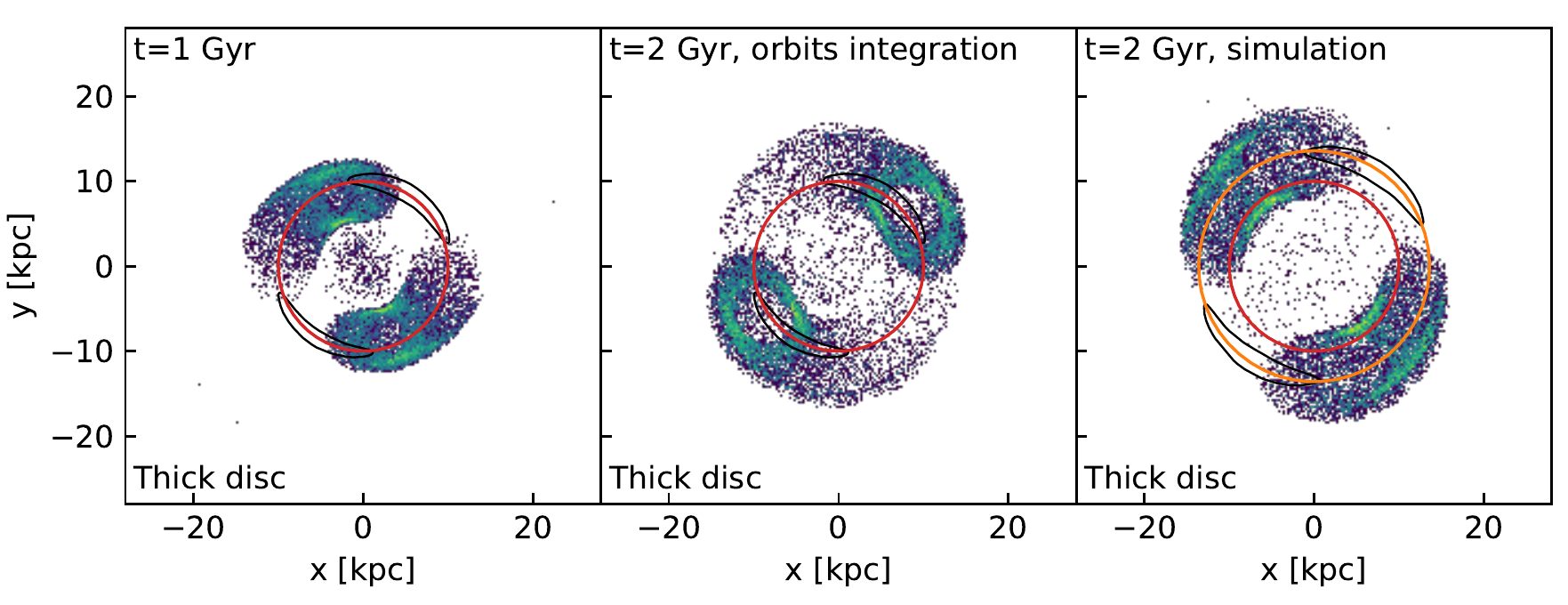} \\
\includegraphics{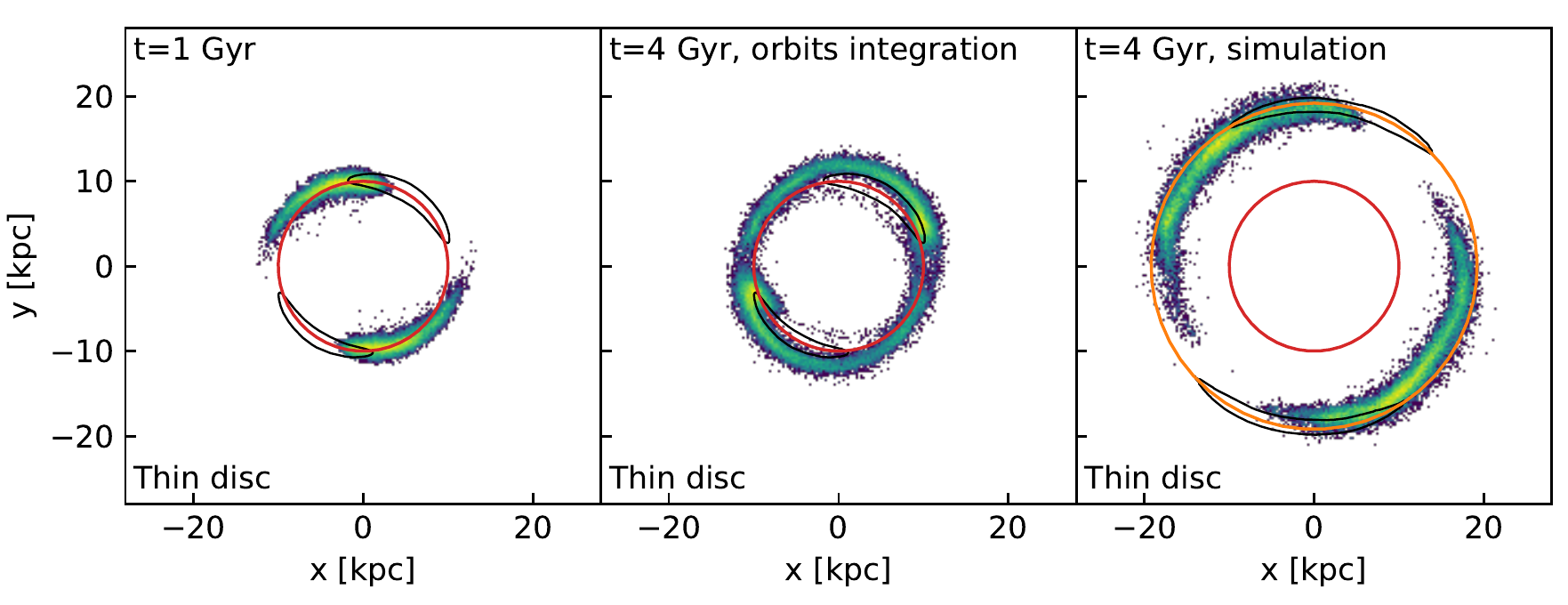} \\
\includegraphics{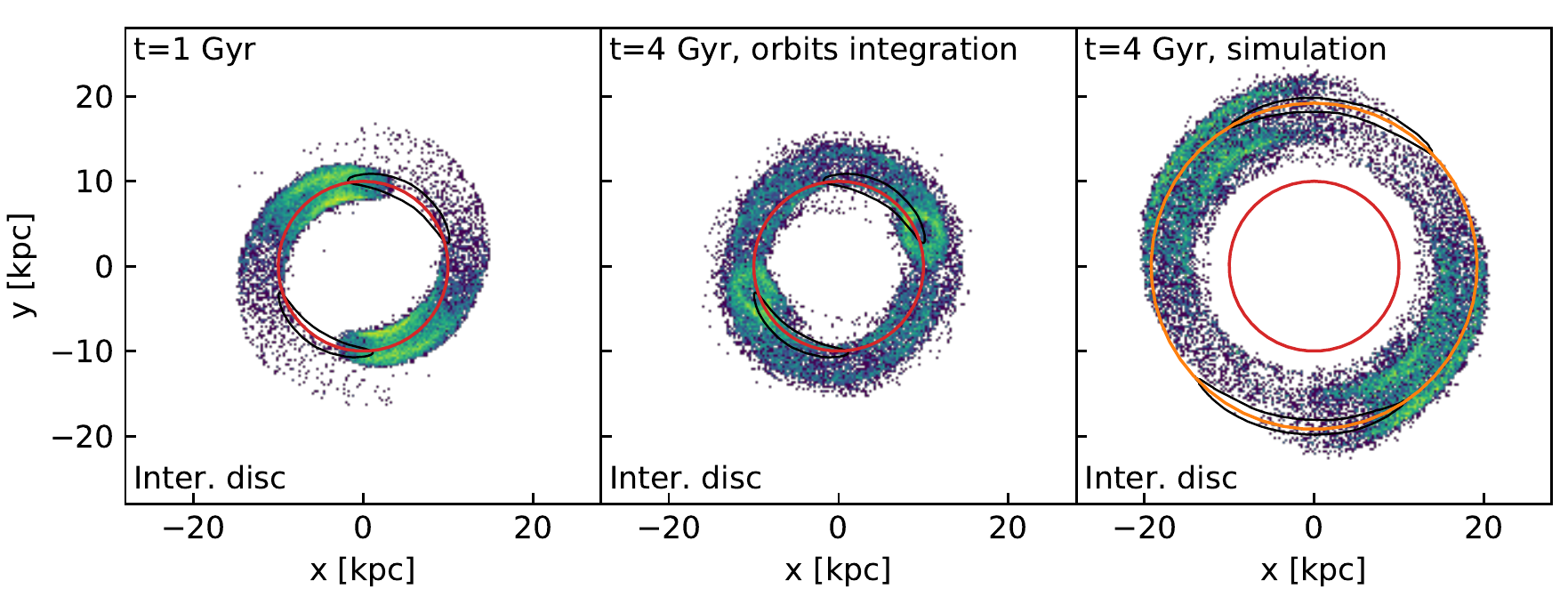} \\
\includegraphics{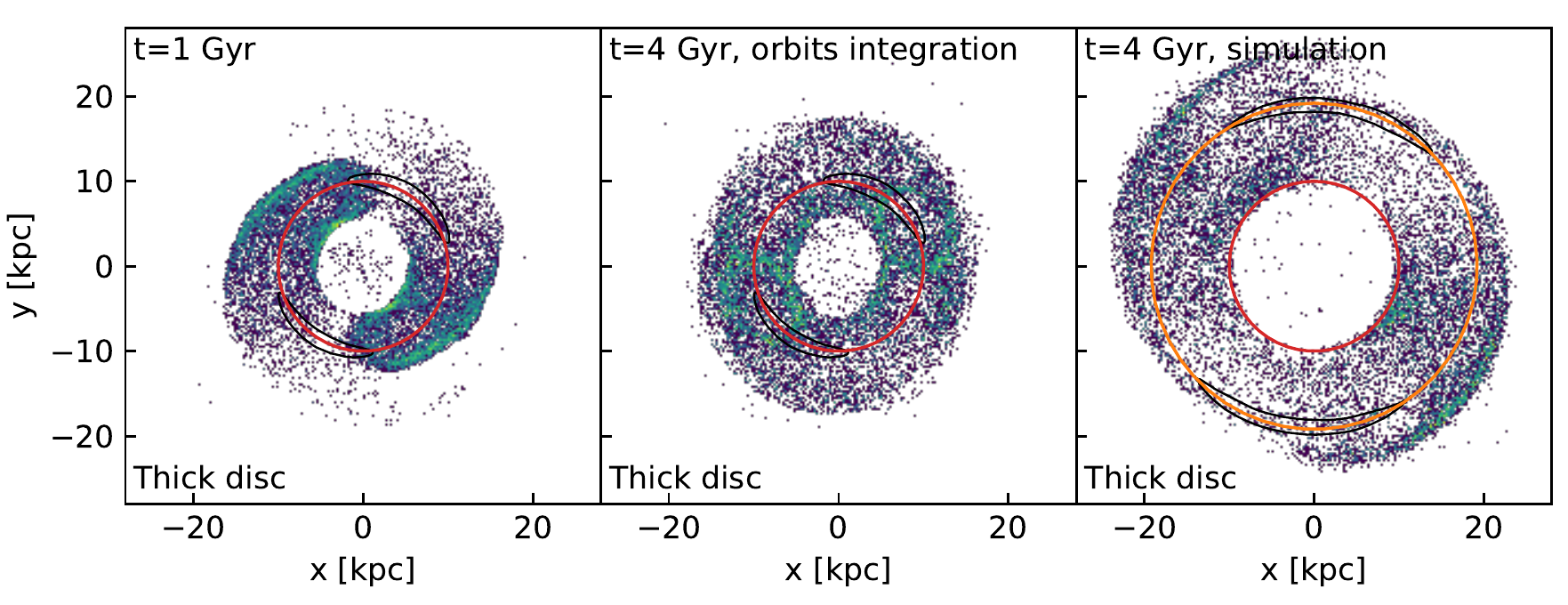} \\
\end{tabular}
}
\caption{Same as Fig~\ref{extrmigrorbs-fig} for the  1$\%$ top stars churned in the simulation.}
\label{extrmigrsim-fig}
\end{figure}

Fig.~\ref{extrmigrorbs-fig} shows the location of the disc stars whose guiding radii increase the most  in the orbits integration (the top $1\%$ of the distribution of change in guiding radius for each disc component) between $t=1$~Gyr and $t=2$~Gyr (top three rows) and between $t=1$~Gyr and $t=4$~Gyr (bottom three rows). Density maps are shown at $t=1$~Gyr, and at the final time ($t=2$ or $t=4$~Gyr) in the orbits integrations case and in the simulation case. At $t=1$~Gyr, the thin disc stars (with low eccentricities and therefore galactocentric radii close to their guiding radii) consist, as expected, of corotating stars in the part of their orbits with a radius lower than corotation. At the final time ($t=2$ or $t=4$~Gyr), they are beyond the (fixed) corotation radius in the orbits integration. Stars of the thicker components are distributed in a wider radial range but their average radius is below the corotation radius at $t=1$~Gyr, and beyond it at the final time.    In the simulation, those stars do not necessarily migrate much: as discussed in the last section, some of them remain at their initial radius while a fraction is churned outwards. Only a very small fraction of stars migrating the most in the orbits integration from $t=1$ to $t=4$~Gyr reaches radii close to the corotation radius at $t=4$~Gyr in the simulation, indicating the extreme migrators in the simulation in this time-interval must consist of different stars.

Fig.~\ref{extrmigrsim-fig} shows the location of the disc stars whose guiding radii increase the most  in the simulation (the top $1\%$ of the distribution of change in guiding radius for each disc component) between $t=1$~Gyr and $t=2$~Gyr (top three rows) and between $t=1$~Gyr and $t=4$~Gyr (bottom three rows). As in Fig.~\ref{extrmigrorbs-fig}, density maps are shown at $t=1$~Gyr, and at the final time ($t=2$ or $t=4$~Gyr) in the orbits integration case and in the simulation case. The spatial distributions at $t=1$~Gyr show that particles that remain trapped at corotation (and are thus significantly churned outwards) are located beyond the local effective potential maxima (in the counter-clockwise rotation sense). Some of these stars are corotating with the bar at $t=1$~Gyr, but the distribution is deprived of stars behind the effective potential local maxima  (in the counter-clockwise rotation sense). The latter stars must be liberated from trapping at corotation, they do not follow the decrease in average angular frequency of the stars trapped at the corotation of a slowing-down non-axisymmetric pattern. The spatial distribution of extreme migrators is even more tilted towards higher azimuths (rotating counter-clockwise) when looking at extreme migrators in the time-interval from $t=1$ to $t=4$~Gyr. Some extreme migrators in the latter time-interval do not corotate with the bar at $t=1$~Gyr (as can be deduced from the spatial distribution of the orbits integration case at $t=4$~Gyr, departing from the spatial distribution of particles at corotation of fig.~\ref{partsatcr-fig}). These stars are trapped by the corotation at a later time (after $t=1$~Gyr).

\begin{figure}[h!]
\centering
\resizebox{\hsize}{!}{\includegraphics{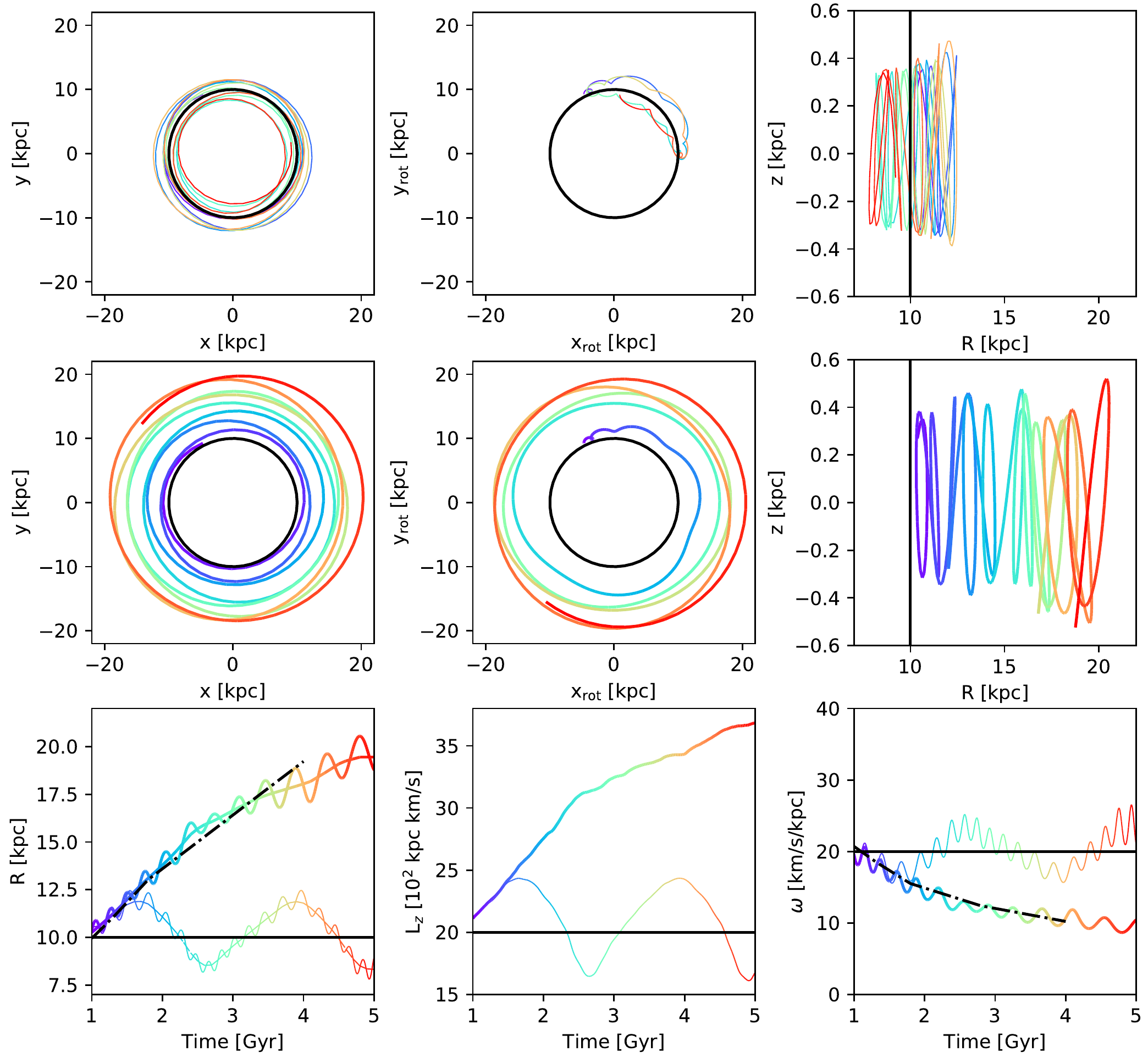}}

\caption{Same as Fig.~\ref{thinorb-fig} for a thin disc stellar particle evolution in the orbits integration (first row and thin lines of third row) and in the simulation (second row and thick lines of third row) in which the stellar particle is an extreme migrator. Black dash-dotted line: time evolution of the corotation radius (bottom-left panel) and of $\Omega_p$ (bottom-right panel) in the simulations.}
\label{thinorbcomp-fig}
\end{figure}

\begin{figure}[h!]
\centering
\resizebox{\hsize}{!}{\includegraphics{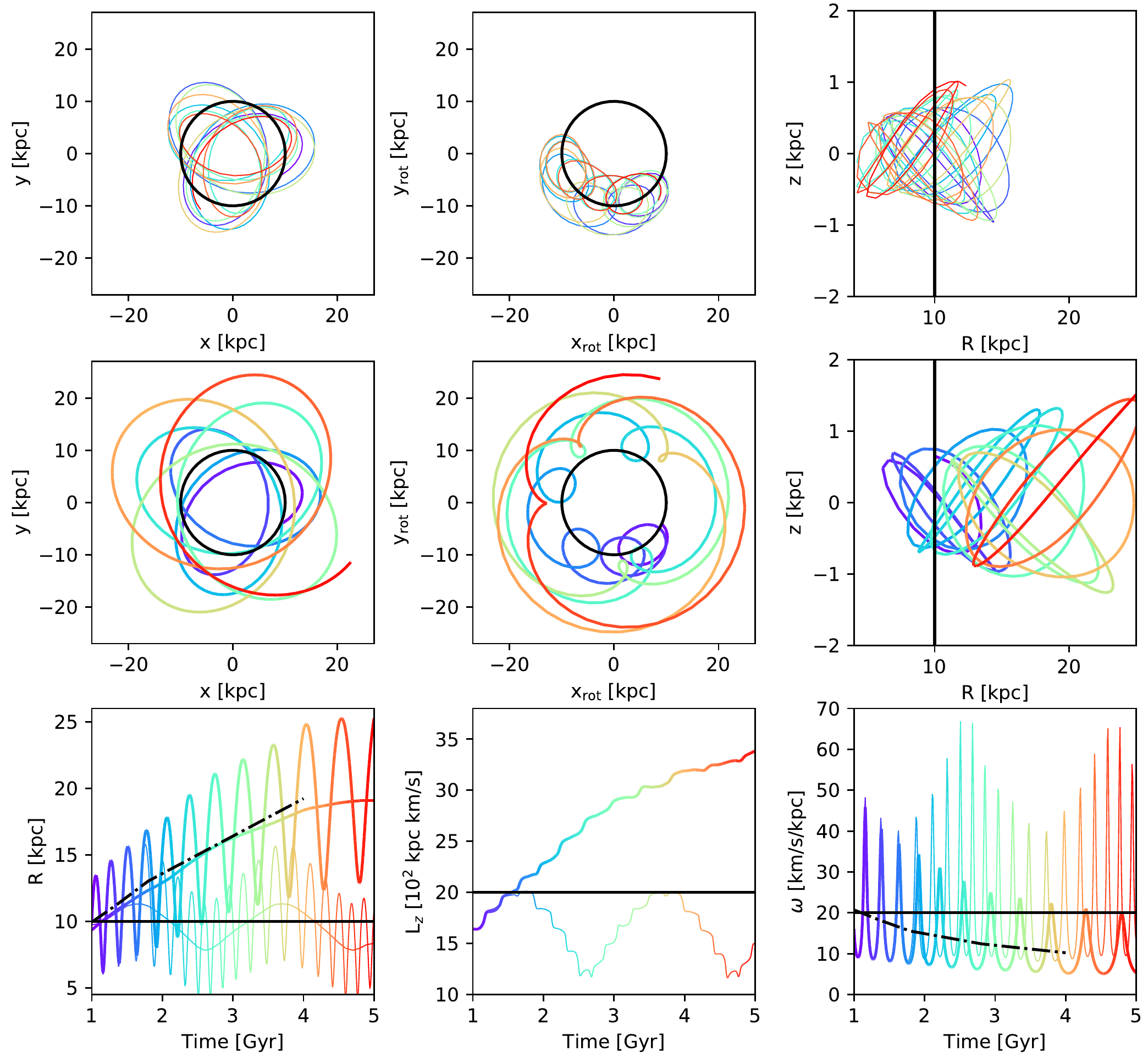}}

\caption{Same as Fig.~\ref{thinorb-fig} for a thick disc stellar particle evolution in the orbits integration (first row and thin lines of third row) and in the simulation (second row and thick lines of third row) in which the stellar particle is an extreme migrator. Black dash-dotted line: time evolution of the corotation radius (bottom-left panel) and of $\Omega_p$ (bottom-right panel) in the simulations.}
\label{thickorbcomp-fig}
\end{figure}

Finally, we compare the orbits of extreme migrators in the simulation to the orbits of the same stars in the orbits integration. Fig.~\ref{thinorbcomp-fig} shows such a comparison for a thin disc star and Fig.~\ref{thickorbcomp-fig} for a thick disc star. Both stars corotate with the bar at $t=1$~Gyr as can be seen in the panel showing the libration of the integrated orbits (thin lines) in the frame rotating at the bar speed at $t=1$~Gyr. As they are trapped at the corotation of the slowing-down pattern, their angular speed decreases and the orbits of the simulation circulate around the centre of the galaxy in this frame. The thick disc star has a high eccentricity but it is however trapped at the bar corotation just as thin disc stars, as found by \citet{binney18}, and has a larger and larger guiding radius. 

\section{Conclusion}

We have studied radial migration in a galactic disc with thick components with a bar and corotating spiral arms as the most prominent non-axisymmetry. The dark matter halo is massive and concentrated, so as to enhance the shifting of resonances radii of interest in this work. 

Stars of the thick components can have large radial excursions because of the high eccentricities of their orbits but the churning is limited and very similar to the churning in the thin disc. This is consistent with results of \citet{solway12} that found thick disc stars undergo almost the same churning as thin disc stars. Note that during the dynamical evolution, the thin disc, initially less stable than the thicker components, heats up, which reduces the difference between the disc components in velocity dispersion.

We have shown that stars belonging to thick components can be trapped at the bar corotation, as found by \citet{binney18}. If the bar keeps the same strength and speed, trapped stars are churned periodically outwards and inwards at corotation, but if the corotation radius increases (in the case of a slowing-down bar), they can be churned to larger guiding radii on average. This outwards churning is possible even in disc components of decreasing total angular momentum because the loss of angular momentum occurring for stars building up the bar makes the global angular momentum decline as time goes by.

Extreme migrators constitute only a fraction of stars at corotation at an initial time, this fraction decreasing as time goes by and being a little smaller for thicker components. These stars can be distinguished from their spatial location with respect to the maxima of the local effective potential extrema at initial time.

Potential observational signatures (in kinematics or metallicity/chemistry) of these extreme migrators of thin and thick components will be studied in a following paper.

\begin{acknowledgements}
PDM and MH acknowledge support from the ANR (Agence Nationale de la Recherche) through the MOD4Gaia project (ANR-15-CE31- 0007, P.I.: P. Di Matteo). This work was granted access to the HPC resources of CINES under the allocation 2016-040507 made by GENCI.
\end{acknowledgements}

\bibliographystyle{aa} 

\bibliography{bibthick}

\begin{appendix}
\section{Rotation curves and velocity dispersions}

\label{rotcurves-app}

\begin{figure}[h!]
\centering
\resizebox{\hsize}{!}{\includegraphics{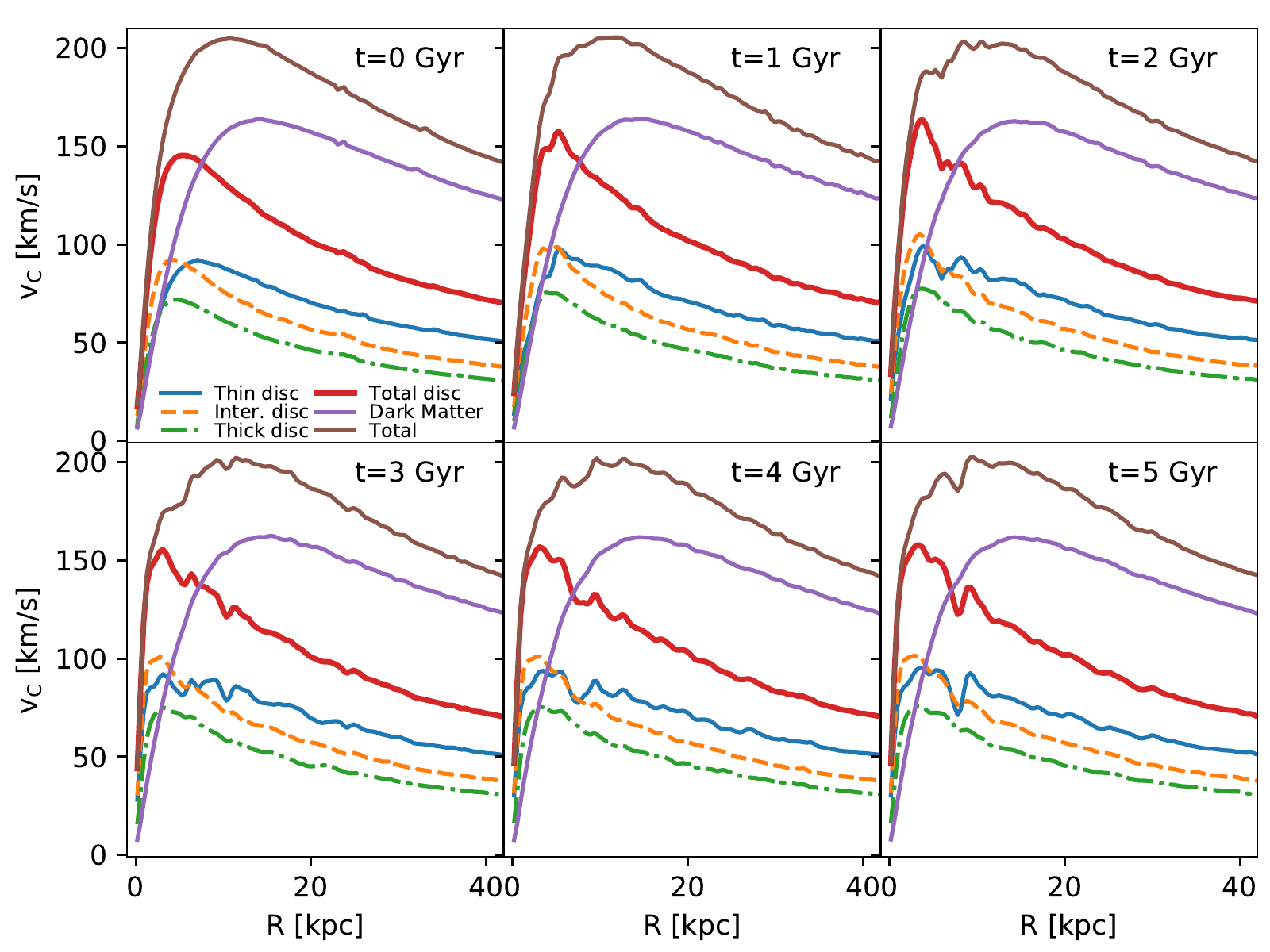} }
\caption{Contribution of the disc and dark matter components to the rotation curve at different times.}
\label{vc-fig}
\end{figure}

\begin{figure}[h!]
\centering
\resizebox{\hsize}{!}{\includegraphics{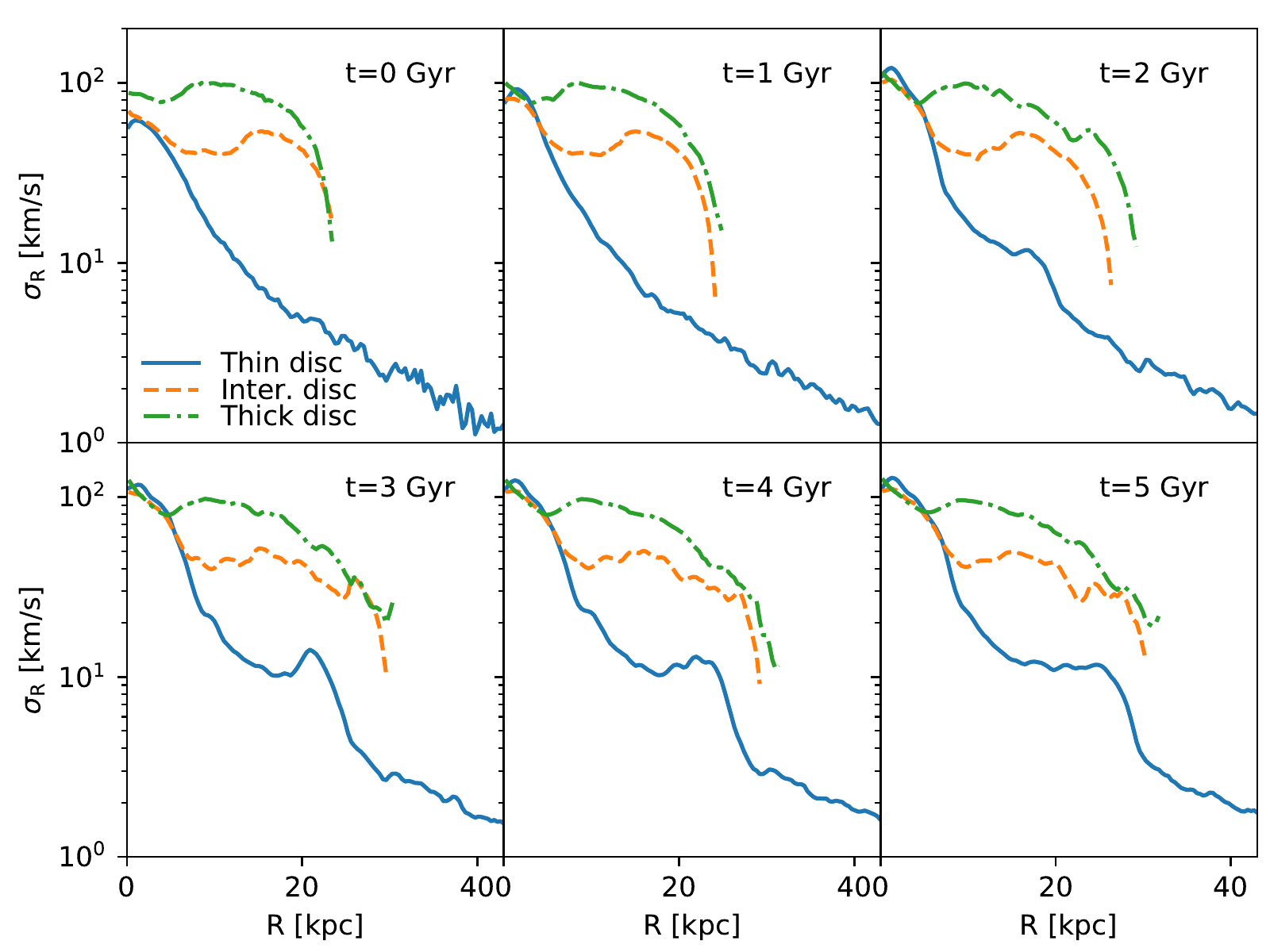} }
\caption{Radial velocity dispersion profiles of disc components at different times.}
\label{dispr-fig}
\end{figure}

In this appendix, we show initial conditions details in addition to section~\ref{subsec-ics}, as well as the evolution of some quantities: rotation curves and radial and vertical velocity dispersions of the disc components. Fig.~\ref{vc-fig} shows the contributions of the different disc components, the whole disc, and the dark matter halo in the initial conditions and at later times. The disc dominates the gravitational potential at low radii at all times, while the dark matter halo dominates at high radii. Figs.~\ref{dispr-fig} shows the radial and vertical velocity dispersions of the different disc components in the initial conditions and at later times.
\begin{figure}[h!]
\centering
\resizebox{\hsize}{!}{\includegraphics{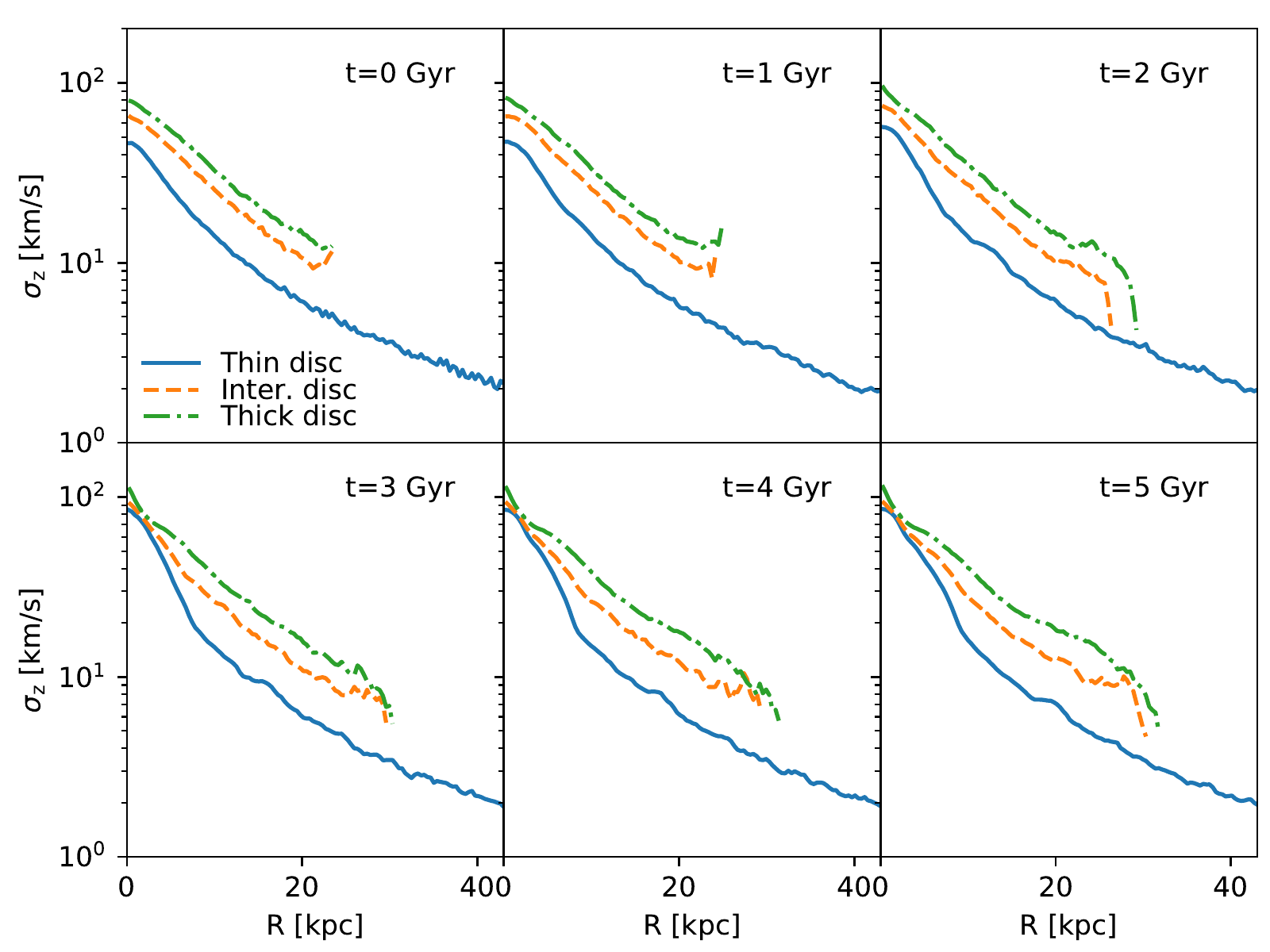}}
\caption{Vertical velocity dispersion profiles of disc components at different times.}
\label{dispz-fig}
\end{figure}

\section{Stars at ILR and OLR}
\label{ilrolr-app}

In this appendix, we show stars found at the ILR and the OLR for the different disc components by the analysis of section~\ref{subsec-crparts}.

\begin{figure*}[h!]
\centering
\begin{tabular}{ccc}
\includegraphics[width=5.5cm]{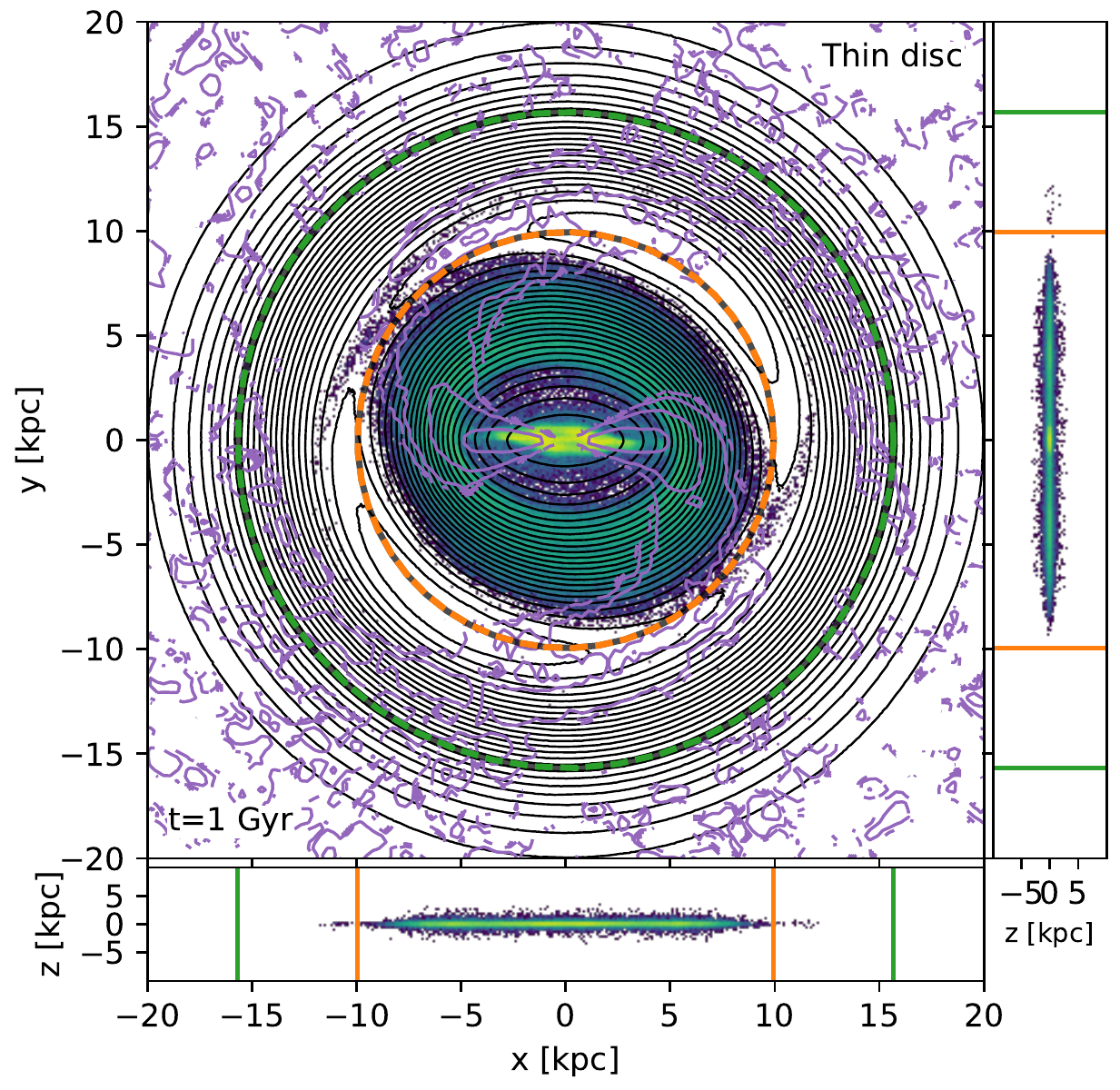} & \includegraphics[width=5.5cm]{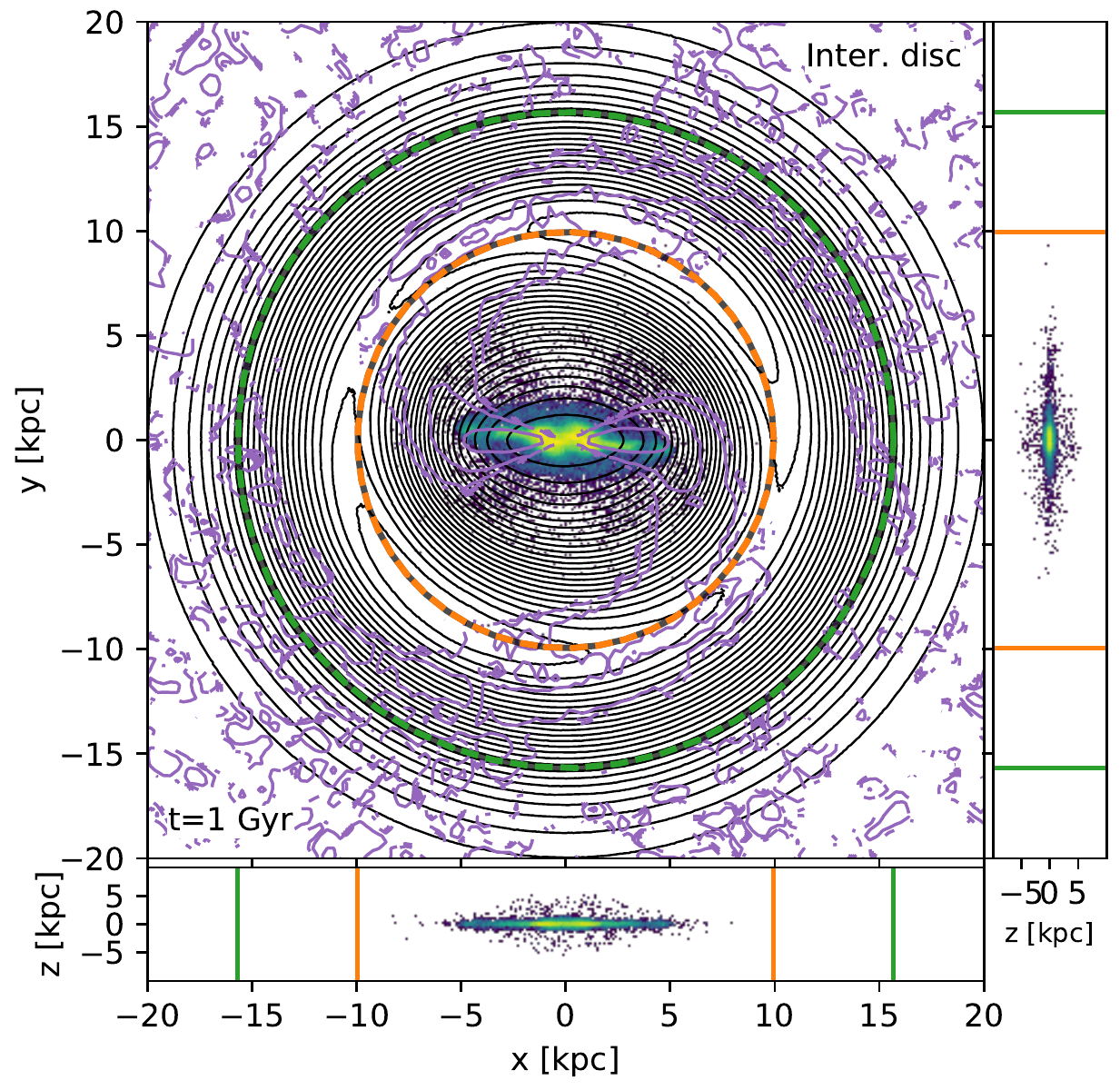} 
& \includegraphics[width=5.5cm]{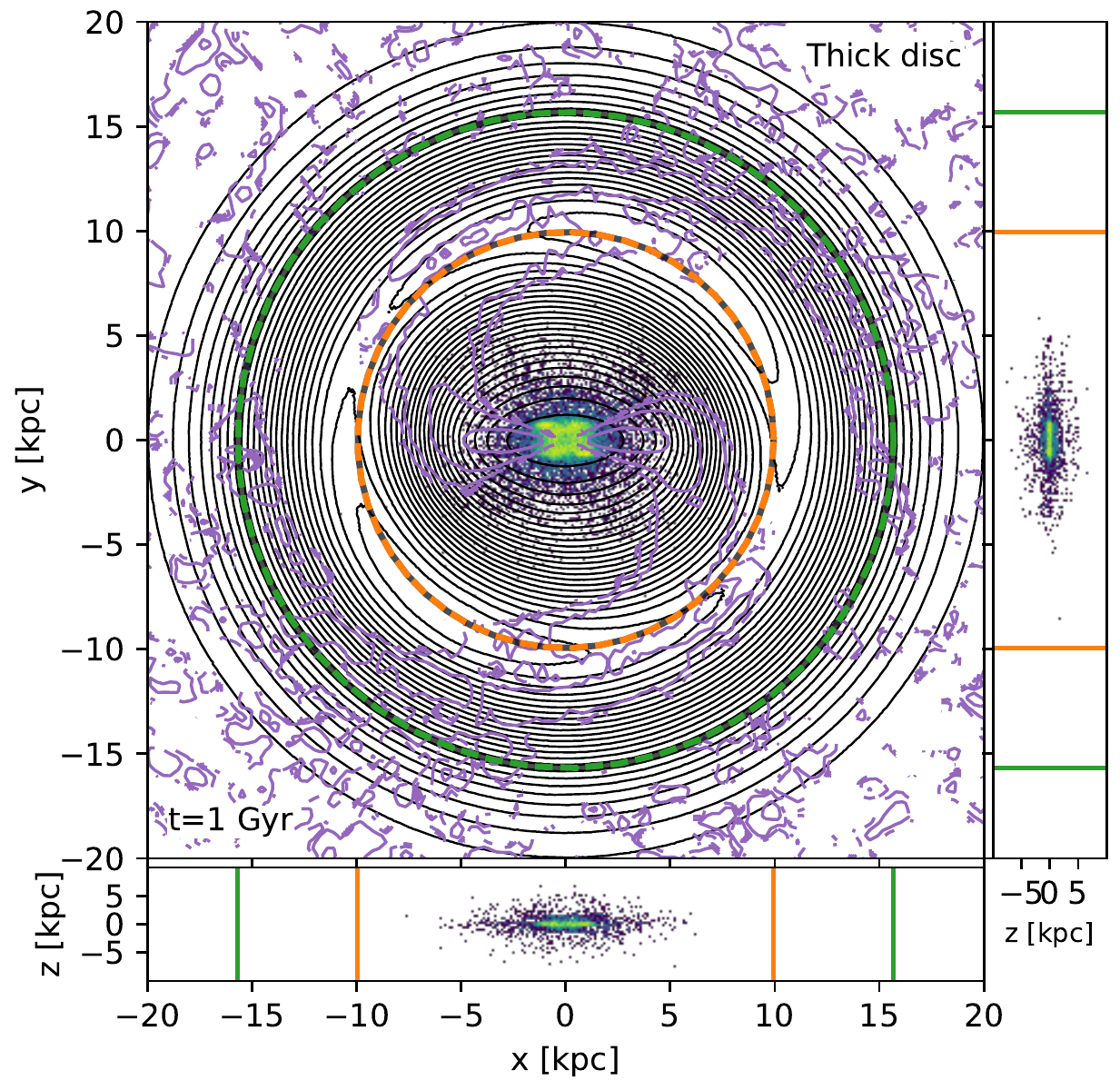} 
\end{tabular}
\caption{Colours: density map of stars at the bar ILR. Black contours: effective potential in a frame rotating at the bar speed. Purple contours: positive azimuthal overdensity in radial bins.}
\label{locilrr-fig}
\end{figure*}

\begin{figure*}[h!]
\centering
\begin{tabular}{ccc}
\includegraphics[width=5.5cm]{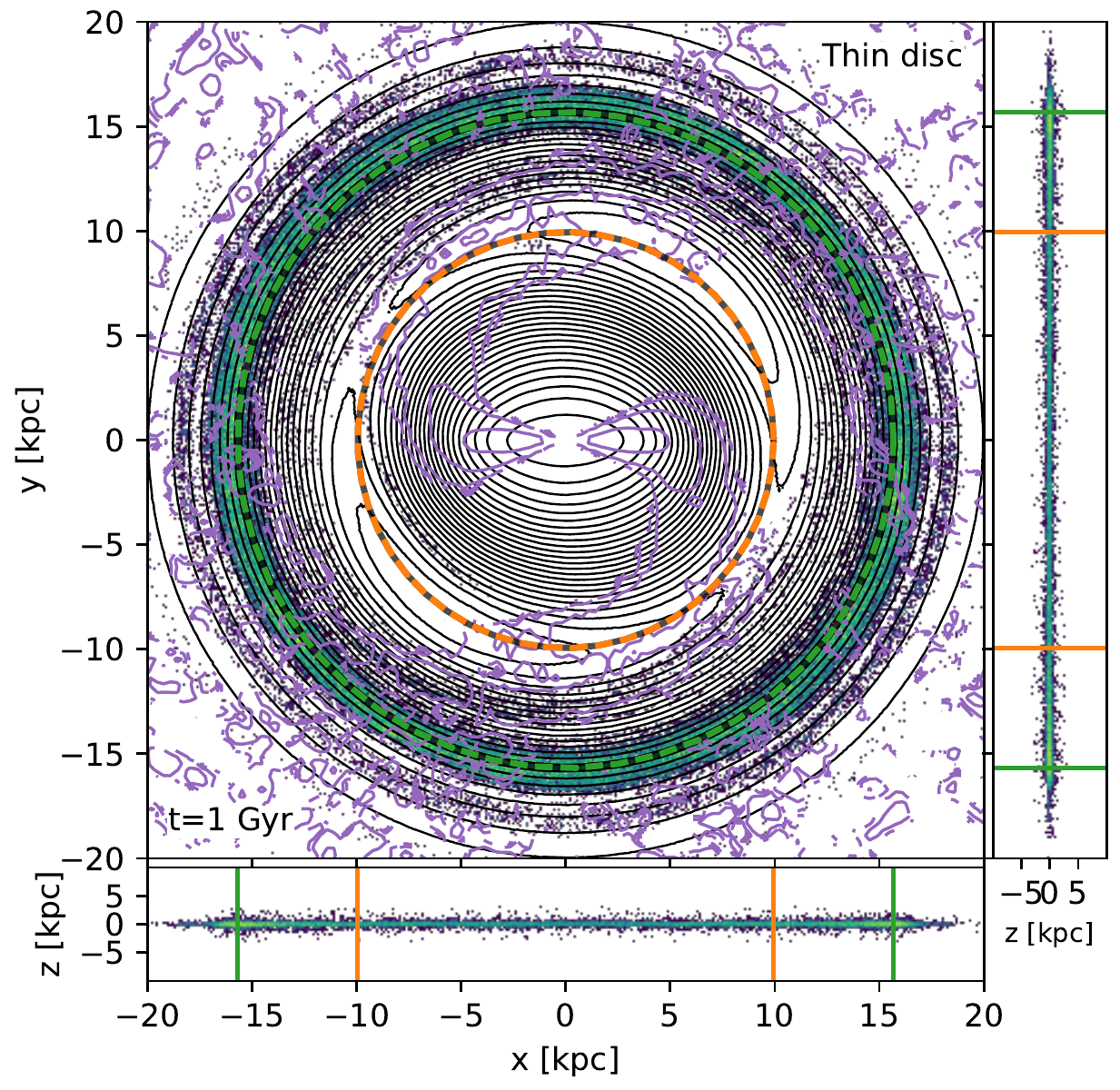} & \includegraphics[width=5.5cm]{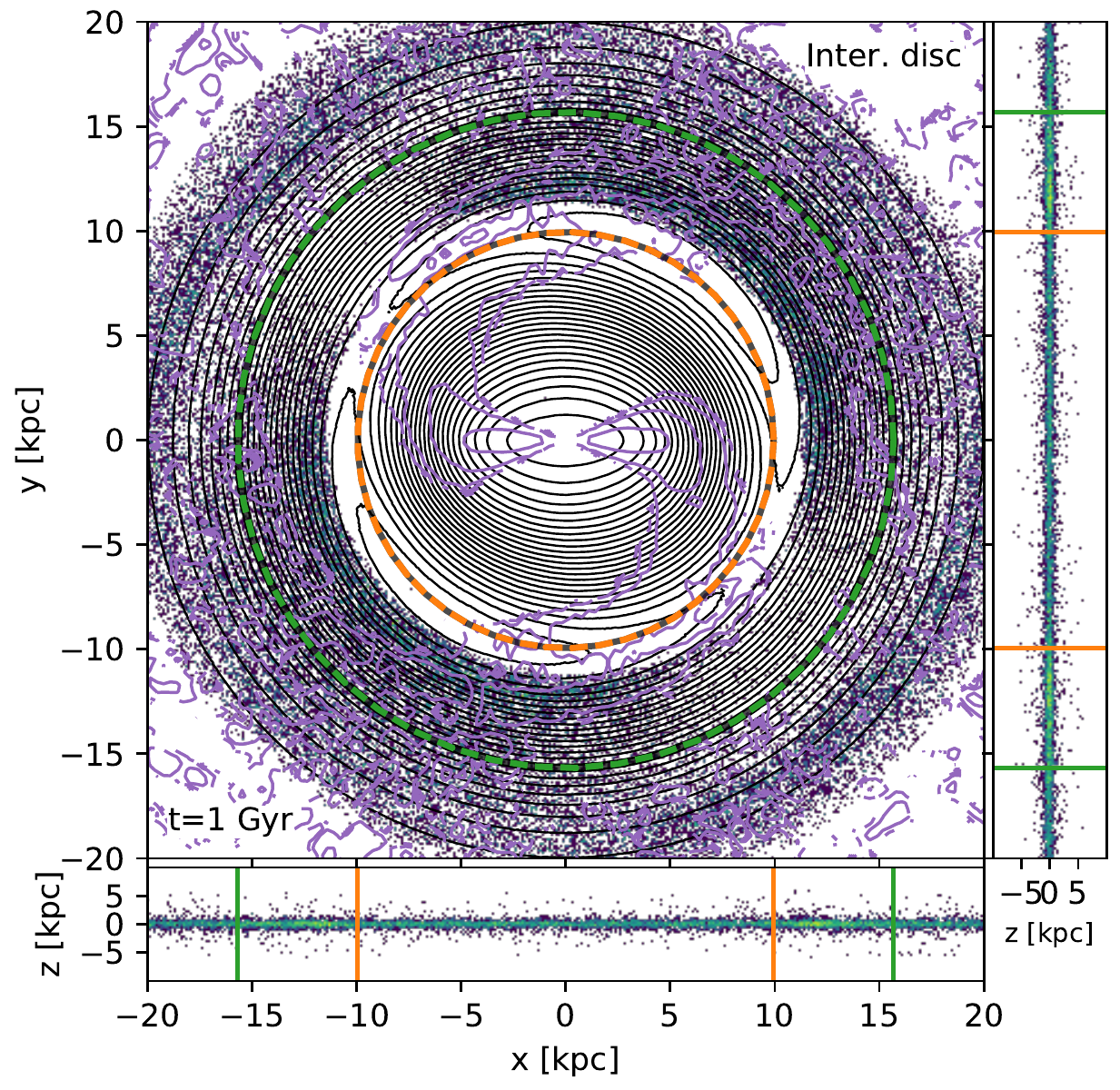} 
& \includegraphics[width=5.5cm]{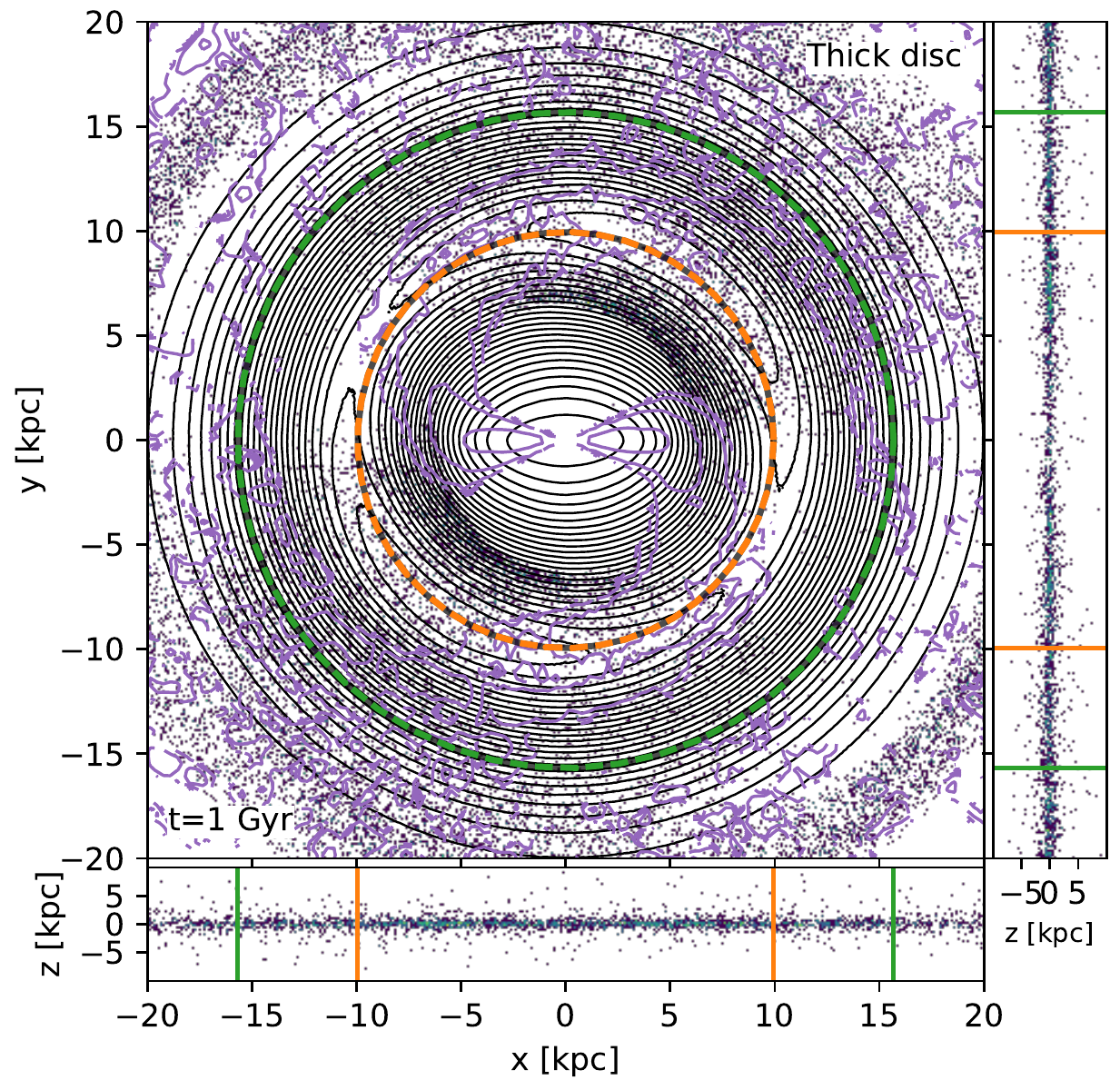} 
\end{tabular}
\caption{Colours: density map of stars at the bar OLR. Black contours: effective potential in a frame rotating at the bar speed. Purple contours: positive azimuthal overdensity in radial bins.}
\label{locolr-fig}
\end{figure*}

Figs.~\ref{locilrr-fig} shows the density maps of stars at the ILR in the different disc components. They mostly belong to the bar. Fig.~\ref{locolr-fig} shows the density maps of stars at the OLR. While the distribution is an annulus for the thin disc stars, it has a wider radial extent for the thick components.

\end{appendix}

\end{document}